\documentclass[preprint,12pt]{elsarticle}
\usepackage{bm}
\usepackage{amssymb}
\journal{Nuclear Physics B}

\begin{document}

\begin{frontmatter}


\title{A new strategy to microscopic modelling of topological entanglement in polymers based on field theory}


\author{Franco Ferrari}
\ead{franco@feynman.fiz.univ.szczecin.pl}

\address{CASA* and Faculty of Mathematics and Physics, University of Szczecin, Wielkopolska 15, 70-451 Szczecin, Poland}

\begin{abstract}
  In this work a new strategy is proposed in order to build analytic and microscopic models of fluctuating polymer rings subjected to topological constraints. The topological invariants used to fix these constraints belong to a wide class of the so-called numerical topological invariants. For each invariant it is possible to derive a field theory that describes the statistical behavior of knotted and linked polymer rings following a straightforward algorithm. The treatment is not limited to the partition function of the system, but it allows also to express the expectation values of general observables as field theory amplitudes.

  Our strategy is illustrated taking as examples the Gauss linking number and a topological invariant belonging to a class of invariants due to Massey. The consistency of the new method developed here is checked by reproducing a previous field theoretical model of two linked polymer rings.
After the passage to field theory, the original topological constraints imposed on the fluctuating paths of the polymers become constraints over the configurations of the topological fields that mediate the interactions of topological origin between the monomers. These constraints involve quantities like the cross-helicity which are of interest in other disciplines, like for instance in modeling the solar magnetic field.

While the calculation of the expectation values of generic observables remains still challenging due to the complexity of the problem of topological entanglement in polymer systems, we succeed here to reduce the evaluation of the moments of the Gauss linking number for two linked polymer rings to the computation of the amplitudes of a free field theory.
\end{abstract}


\end{frontmatter}


\section{Introduction}
\label{Introduction}
Topological structures appear everywhere in nature and shape the properties of physical systems at practically almost every scale of length, from proteins and long polymers to stars. Eliminating knots and links in polymer materials allows to construct polymer networks that are supersoft and superelastic \cite{PaRu}. On the other side of the scale, it is known that topology constrains the energy stored in the solar magnetic field \cite{Woltjer1,Woltjer2}. It is important to quantify the topological complexity of the solar magnetic lines in order to predict phenomena that are potentially dangerous for our civilization like solar flares and coronal mass ejections \cite{Schmieder,Priest, Janvier}.

One of the most outstanding tasks of polymer physics is to construct analytic and microscopic models that are able to characterize the behavior of matter in the presence of topological constraints. The main difficulty is that in physics the interactions are traditionally taken into account by potentials, but the entanglement of long, quasi one-dimensional objects cannot be easily described in this way. Some help in the case of polymers with closed conformations has come from the advances made by knot theory in the last decades. Nowadays powerful topological invariants are available in order to distinguish the different topological states of a given system. Field theoretical models of topological entanglement based on some of these invariants already exist. Their main feature is that topological field theories are responsible for mediating the interactions between the monomers arising due to the presence of the topological constraints. The first field theory describing the statistical behavior of two rings  in a solution and linked together has been derived in Ref.~\cite{FFIL1998}. Before, several other models have been proposed, in which the fluctuations of one of the polymers have been approximated in several ways, for instance by considering them as white noise \cite{Edwards,Vilgis,VilgisOtto,Otto,Nechaev,Kholodenko,Brereton}. In \cite{FFIL1998} the topology of the two-ring system is fixed using the Gauss linking number, a topological invariant that is related to the vacuum expectation values of the observables of a topological field theory called the Abelian BF model \cite{BlauThompson1991}.
 The vector bonds of the two polymer rings are replaced by monomer densities, whose fluctuations are taken into account
in the partition function by performing a statistical sum over a set of replica complex scalar fields. 
The interactions due to the presence of the topological constraints are propagated by
the topological fields of the Abelian BF model.
This simple model has allowed some useful predictions. For example, it has been possible to conclude that the topological complexity of a link formed by two polymer rings, measured according to increasing values of the Gauss linking number, rises linearly with the length of the rings \cite{FFHKILPLA2000}. This prediction has been later confirmed by an independent calculation in \cite{Otto}. Moreover, the renormalization group analysis performed in \cite{FFIL1999} has shown that the monomers of two very long polymers entangled together are subjected to attractive interactions of entropic origin due to the presence of the topological constraints. This attraction has also been observed experimentally in \cite{Quake}. On the other side, it is known that unentangled polymer rings slightly repel each other even in the absence of excluded volume forces. In \cite{FF2004} the co-existence of attractive and repulsive forces in a system of two linked polymer rings has been proved with the help of nonperturbative methods.

Once a model has been derived for a given topological system, it can be useful to understand at least some of the features of other systems
in which topological relations play an important role. Indeed, topological matter may be realized using very different quasi one-dimensional objects, like for instance polymers or the  lines in the solar magnetic field, but the behavior of these objects is influenced by the same topological constraints. For instance, the Abelian topological field theory \cite{FFIL1998}, originally derived for polymers, can be mapped into a theory of quasi-particles in multi-layered superconductors \cite{FF2004} and is relevant for topological computing too \cite{dassarma,nayak,wilczek,goldman,anyonexp2013,gurarie,dolev,field-simula}. On the other side, the topological field theories of \cite{nova,non-semisimple}, constructed having in mind the application to polymer physics of more powerful topological invariants than the Gauss linking number, turn out to be important for topological insulators \cite{Juven,He,McLeish,Chen}.%3.05.2019 References written up to this point

The major drawback of the Gauss linking number is that it is a weak topological invariant. It is not uncommon that the values of the Gauss linking number computed in the case of two topologically inequivalent links are the same, so that the two links cannot be distinguished according to this invariant.
%An approach called topological engineering that allows to construct models exploiting more refined invariants has %been presented in \cite{nova}.
In order to construct models based on more refined invariants it is possible to follow different strategies.
Here we choose the approach pioneered by Sam Edwards~\cite{Edwards} which has shown to be very promising in the past.
This approach makes use of the so-called numerical topological invariants. The latter have the advantage that can be expressed in the form of contour integrals computed along the paths of the loops composing a knot or link. As a consequence, these invariants depend explicitly on the physical paths of the polymers. The topological constraints are imposed on a set of $N$ loops $C_1,\ldots, C_N$  by requiring that the value of a given numerical topological invariant $\tau_N(C_1,\ldots,C_N)$ should be equal to some number $m$. The statistical sum over all loop conformations is performed via path integral techniques~\cite{Kleinertbook}. The constraint $\tau_N(C_1,\ldots,C_N)-m=0$ is taken into account by inserting in the partition function a Dirac delta function enforcing that condition. The presence of a Dirac delta function in a path integral is quite ackward. This shortcoming is circumvented exploiting the Fourier representation $\delta(\tau_N-m)=\int_{-\infty}^{+\infty}\frac{d\lambda}{\sqrt{2\pi}} e^{-i\lambda(\tau_N-m)}$. In this way a new parameter $\lambda$ is introduced, namely the Fourier variable conjugate to $m$.
Yet, the expression of $\tau_N(C_1,\ldots,C_N)$ consists in general of very complicated multiple contour integrals over the paths $C_1,\ldots,C_N$, a fact that makes the evaluation of the statistical sum over the loops $C_1,\ldots,C_N$ an outstanding problem.
It is at this point that field theory comes into play. The idea of \cite{nova} is
to construct a topological field theory with the special characteristics that: i) it is possible to define $N$ observables ${\cal O}_1,\ldots,{\cal O}_N$ such that their expectation value  $\langle {\cal O}_1\ldots,{\cal O_N}\rangle$ coincides with $e^{-i\lambda \tau_N(C_1,\ldots,C_N)}$; ii) The observables ${\cal O}_1,\ldots,{\cal O}_N$ are non-local objects similar to the Wilson loops possibly with some slight modifications in order to make them both gauge invariant  and metric independent; iii) the whole dependence on the $i-$th loop $C_i$, where $i=1,\ldots,N$, is confined to the $i-$th observable, i.~e.: ${\cal O}_i={\cal O}_i(C_i)$.  If such a theory exists, the path integration over all possible loop conformations can be split into $N$ separate integrations in which only a single loop is considered. This is a consistent simplification that, together with property ii), allows
a straightforward passage from paths to monomer densities using standard second quantization techniques \cite{nova}.
As an upshot, the  partition function of a system of interlocked polymer rings is mapped into the partition function of a field theory that is both polynomial and local, thus eliminating the non-polynomiality and non-locality that were present in the original formulation.
In the case of the Gauss linking number, an example of topological field theory that satisfies the requirements i)--iii) is the already mentioned Abelian BF model of Ref.~\cite{FFIL1998}.
For higher order invariants, special Chern-Simons field theories with non semi-simple gauge groups are necessary \cite{non-semisimple,nova}. Numerical topological invariants are often characterized by path ordering, so that the order of the integrations in the multiple contour integrals over the loop paths is constrained. In this case Chern-Simons fields must be coupled with auxiliary one-dimensional fields that take into account the path ordering.
A topological field theory of this kind related to a topological invariant called the triple Milnor invariant has been derived in \cite{FFMPYZ2015}.

Despite many efforts and some relevant advances that have been achieved with the help of field theories or alternative methods \cite{Nechaev,Kholodenko,Rohwer}, the modeling of topological structures in polymer systems is still affected by several limitations.
In what concerns the
techniques based on field theory worked out in \cite{FFIL1998,non-semisimple,nova,FFMPYZ2015}, their validity is restricted to a few invariants and, in the case of an arbitrary invariant, there is no general algorithm for building topological field theories satisfying the properties i)--iii) defined above.
It is lecit to expect that the difficulties of deriving such field theories will increase with increasing complexity of the topological invariant used to fix the constraints. In the case of knots, even the simplest numerical topological invariant is characterized by a very complex mathematical expression. This is  the major reason for which up to now no microscopic analytical model exists that is able to describe the statistical behavior of a single knotted polymer ring.
Moreover, so far only the partition function of the interlocked polymer rings can be mapped to field theory.
No generalization of this mapping exists if, instead of the partition function, the averages of observables like the gyration radius or the distance between two points of the link are considered.
A further complication stems out from the fact that the coupling constant that determines in the field model the strength of the interactions between the monomers due to the topological constraints is the Fourier parameter $\lambda$.
This coupling constant can take any possible value on the real line, making it difficult to choose a suitable approximation for performing any calculation.

In this paper a new approach to construct field theoretical models of links formed by polymers and overcome at least in part the above difficulties is presented.
In the previous approaches the lack of potentials describing topologically entangled polymer rings has been solved by the introduction of very special topological field theories that are able to reproduce with their amplitudes the expression of the invariant used to fix the topological constraints. On one side this strategy is very elegant, because it considers the interactions of topological origin between the monomers in the same way as the fundamental interactions in high energy physics. The only difference is that in polymer systems the interactions originate from the presence of topological constraints and are mediated by topological fields. On the other side, the models obtained in this way are affected by the drawbacks explained before. With the present work potentials gain back the leading role. The idea is that numerical topological invariants can be regarded as potentials.
Despite the fact that the expressions of such invariants are complicated, being both non-local and non-polynomial, it will be shown here that it is possible to simplify them with the help of field theories. The starting point is the observation that numerical topological invariants consist of a sum of multiple contour integrals along the paths of the $N$ loops composing a given link.  The loops are curves in space whose positions are specified by the position vectors $\boldsymbol r_1(s_1),\ldots,\boldsymbol r_N(s_N)$, where the parameters $s_1,\ldots,s_N$ are the arc-lengths  parametrizing the curves.
The trick, which is well known in polymer physics and  in solar magnetohydrodynamics, is to replace the position vectors  by magnetic flux densities concentrated inside the loops. 
Instead of searching for topological field theories whose amplitudes coincide with the exponent of the topological invariant $e^{-i\lambda\tau_N(C_1,\ldots,C_N)}$, $N$ magnetic flux densities $\boldsymbol b_1(\boldsymbol x), \ldots,\boldsymbol b_N(\boldsymbol x)$ are introduced. Suitable constraints relate these densities to the paths of the $N$ loops and allow to write the invariant $\tau_N(C_1,\ldots,C_N)$ as a function of $\boldsymbol b_1(\boldsymbol x), \ldots,\boldsymbol b_N(\boldsymbol x)$: $\tau_N(C_1,\ldots,C_N)=\tau_N(\boldsymbol b_1(\boldsymbol x), \ldots,\boldsymbol b_N(\boldsymbol x))$. This procedure, which consists in a few simple steps that will be stated  precisely in the Conclusions, is sufficient in order to disentangle the loops, reducing the statistical sum over all their possible conformations to $N$ independent path integrals, each one involving a single loop. Using standard techniques of second quantization,  for each topological invariant a field theoretical model is obtained, in which the action governing the fields is both polynomial and local. Surprisingly, this simple strategy has never been investigated in the over fifty years of history of the application of field theory to build models of topologically entangled polymers.

The new way of mapping the problem of topologically entangled polymer into a field theory presents several advantages with respect to the previous ones:
\begin{enumerate}
\item In the final field theoretical model the connection with the original topological constraints is not lost. The requirement $\tau_N(C_1,\ldots,C_N)=m$ is substituted by the following condition on the magnetic flux densities: $\tau_N(\boldsymbol b_1,\ldots,\boldsymbol b_N)=m$ \label{aaa}.
\item The passage from the statistical sum over the loop conformations to a path integral over fields is straightforward and can be easily generalized to any numerical topological invariant.
\item There is no need as in previous approaches to use the Fourier representation of Dirac delta function imposing the topological constraints. 
We will show here that in this way the expression of the second topological moment of the Gauss linking number, that could be  calculated only approximately in Ref.~\cite{FFHKILPLA2000}, may be derived in closed form after evaluating the amplitudes of a free field theory.
\item The passage from loops to fields may be easily extended to include also the expectation values of the observables and not only the partition function of the entangled polymer system.
\end{enumerate}
In this work the new method is applied to numerical topological invariants with no path ordering. First, the procedure is tested in the case of the Gauss linking number denoted here with the symbol $\tau_2(C_1,C_2)$. The final field theory is a Ginzburg-Landau type model with replica scalar fields coupled to abelian BF fields.
The equivalence of  this model with the previous model of Ref.~\cite{FFIL1998} is proved.
The original condition on the Gauss linking number $\tau_2(C_1,C_2)=m$ is replaced by the condition that the so-called cross-helicity of two abelian BF fields is equal to $m$. The cross-helicity is a well known topological invariant used for instance in solar magnetohydrodynamics \cite{Moffatt,Arnold,Hornig}. Next, we build a field theory model corresponding to a third order topological invariant that has been derived in \cite{non-semisimple,nova} by computing the amplitudes of the observables of a topological field theory with a non semi-simple gauge group of symmetry. The same invariant has also been proposed to describe the topological entanglement of the lines of the solar magnetic field, see \cite{Hornig}. With respect to the previous methods, not only the partition function of the system, but also the averages of a general class of observables are mapped into the amplitudes of replica scalar fields.   
\section{The case of the Gauss linking number}
The Gauss linking number $\tau_2(C_1,C_2)$ is the simplest example of numerical topological invariants.
It describes the topological states
of two closed polymers whose paths $C_1$ and $C_2$ are linked together. To simplify the notations, we will assume that both loops have the same length $L$.
Representing $C_1$ and $C_2$ as curves parametrized by the arc-lengths $s_1$ and $s_2$ respectively, with $0\le s_1,s_2\le L$, it is possible to express $\tau_2(C_1,C_2)$ in a form in which the dependence on the physical conformations $C_1$ and $C_2$ is explicit:
\begin{equation}
\tau_2(C_1,C_2)=\frac 1{4\pi}\int_0^Lds_1\int_0^Lds_2\boldsymbol {\dot r}_1(s_1)\cdot\left[
  \frac{\boldsymbol {\dot r}_2(s_2)\times(\boldsymbol r_2(s_2)-\boldsymbol r_1(s_1))}{\left|
\boldsymbol r_2(s_2)-\boldsymbol r_1(s_1)
    \right|^3}
\right]\label{glnexplicit}
\end{equation}
In Eq.~(\ref{glnexplicit}) the position vectors $\boldsymbol r_1(s_1)$ and $\boldsymbol r_2(s_2) $ specify the positions of the points of $C_1$ and $C_2$ in space and $\boldsymbol{\dot r}_i(s)=\frac{d\boldsymbol r_i(s)}{ds}$.
In the partition function ${\cal Z}[m]$ describing the fluctuations of the two polymer loops the topological states will be restricted by requiring that:
\begin{equation}
\tau_2(C_1,C_2)=m\label{glndef}
\end{equation}
Using the path integral formulation of the statistical mechanics of polymers due to Edwards~\cite{Edwards}, ${\cal Z}[m]$ may be written as follows:
\begin{equation}
  {\cal Z}[m]=\int{\cal D}\boldsymbol r_1(s){\cal D}\boldsymbol r_2(s) e^{
  -S_{pol}^2}
  \delta\left(\tau_2(C_1,C_2)-m\right)\label{partfundef}
\end{equation}
In the above equation we have put
\begin{equation}
    S_{pol}^N=\sum_{i=1}^3\int_0^Lds
  \frac3{2a}
\boldsymbol {\dot r}_i^2(s)
    \label{spol}
\end{equation}
with $N=2$. $a$ denotes the Kuhn length.
Moreover, we have assumed that the path integrations in ${\cal D}\boldsymbol r_i(s)$, $i=1,2$, are performed over loops that start and end in a fixed point $\boldsymbol r_{i,0}$. An additional integration over $\boldsymbol r_{1,0}$ and $\boldsymbol r_{2,0}$ is necessary in order to recover the case of two freely moving polymer rings.

In order to simplify the path integration over $\boldsymbol r_1(s)$ and $\boldsymbol r_2(s)$, we introduce
the new quantity:
\begin{equation}
\tau_2(\boldsymbol b^1,\boldsymbol b^2)=\frac 1{4\pi}\int d^3x\int d^3y \epsilon^{\mu\nu\rho}b_\mu^1(\boldsymbol x)\frac{(x-y)_\nu}{|\boldsymbol x-\boldsymbol y|^3} b^2_\rho(\boldsymbol y)\label{chidblintspat}
\end{equation}
$\boldsymbol b^1(\boldsymbol x)$ and $\boldsymbol b^2(\boldsymbol x)$ being two fictitious magnetic flux densities with spatial components $b_\mu^1(\boldsymbol x)$ and $b_\mu^2(\boldsymbol x)$, $\mu=1,2,3$.
In Eq.~(\ref{chidblintspat}) the symbol $\epsilon^{\mu\nu\rho}$ denotes the completely anti-symmetric tensor in three dimensions defined by the condition $\epsilon^{123}=1$. The identity
\begin{equation}
\tau_2(\boldsymbol b^1,\boldsymbol b^2)=\tau_2(C_1,C_2)
\end{equation}
is satisfied provided $\boldsymbol b^1$ and $\boldsymbol b^2$ are defined as follows:
\begin{equation}
\boldsymbol b^1(\boldsymbol x)=\int_0^Lds\boldsymbol{\dot r}_1(s)\delta(\boldsymbol x-\boldsymbol r_1(s))\label{bxdef}
\end{equation}
\begin{equation}
  \boldsymbol b^2(\boldsymbol y)=\int_0^Lds
  \boldsymbol{\dot r}_2(s)
  \delta(\boldsymbol y-\boldsymbol r_2(s))\label{bydef}
\end{equation}
$\boldsymbol b^1(\boldsymbol x)$ and $\boldsymbol b^2(\boldsymbol x)$ may be thought as magnetic flux densities  generated by  two fictitious currents $\boldsymbol j^1(\boldsymbol x)$ and $\boldsymbol j^2(\boldsymbol x)$
circulating respectively along the loops $C_1$ and $C_2$:
\begin{equation}
\boldsymbol j^i(\boldsymbol x)=\int_0^L ds \boldsymbol{\dot r}_i(s)\frac 1{|\boldsymbol x-\boldsymbol r_i(s)|}\qquad\qquad i=1,2
\end{equation}
Let us notice that, due to the presence of the Dirac delta functions in their definitions, $\boldsymbol b^1$ and $\boldsymbol b^2$ transform under general diffeomorphisms as vector densities and not as pure vectors. This means that, with respect to pure vectors, they pickup after a general coordinate transformation an additional factor consisting of the inverse Jacobian of the transformation. Moreover, it turns out from the definitions of Eqs.~(\ref{bxdef})--(\ref{bydef}) that both  $\boldsymbol b^1$ and $\boldsymbol b^2$ are purely transversal, i.~e.
\begin{equation}
  \boldsymbol\nabla\cdot\boldsymbol b^{1,2}(\boldsymbol x)=0\label{condforgaugeinv}
\end{equation}
This is because the line integral around a closed loop of the gradient of a function that is not multivalued is always zero. 
Taking into account Eqs.~(\ref{chidblintspat})--(\ref{bydef}), the partition function of Eq.~(\ref{partfundef}) becomes:
\begin{eqnarray}
  {\cal Z}[m]&=&\int{\cal D}\boldsymbol r_1(s){\cal D}\boldsymbol r_2(s){\cal D}\boldsymbol b^1(\boldsymbol x){\cal D}\boldsymbol b^2(\boldsymbol x) e^{-S_{pol}}\nonumber\\
  &\times&\delta\left(\frac 1{4\pi}\int d^3x\int d^3y \epsilon^{\mu\nu\rho}b_\mu^1(\boldsymbol x)\frac{(x-y)_\nu}{|\boldsymbol x-\boldsymbol y|^3} b^2_\rho(\boldsymbol y)-m\right)\nonumber\\
  &\times&\prod_{i=1}^2\delta\left(
  \boldsymbol b^i(\boldsymbol x)-\int_0^Lds\boldsymbol{\dot r}_i(s)\delta(\boldsymbol x-\boldsymbol r_i(s))
  \right )\prod_{j=1}^2\delta\left(
\boldsymbol \nabla\cdot\boldsymbol b^j(\boldsymbol x)
  \right)\label{zpartfunthree}
\end{eqnarray}
In the last line of the above equation the product of the two functional delta functions enforces the conditions (\ref{bxdef})--(\ref{bydef}) on the fields $\boldsymbol b^{1,2}$. These conditions affect only the tranverse degrees of freedom of $\boldsymbol b^1$ and $\boldsymbol b^2$.
For this reason, 
the additional product of delta functions  $\prod_{j=1}^2\delta\left(
\boldsymbol \nabla\cdot\boldsymbol b^j(\boldsymbol x)
  \right)
  $ has been inserted in the partition function in order  to impose also the transversality conditions of Eq.~(\ref{condforgaugeinv}).
Technically speaking, ${\cal Z}[m]$ is the partition function of an abelian gauge field theory quantized in the pure Lorentz gauge. For such kind of theories the longitudinal degrees of freedom are present only in the gauge fixing terms.

At this point we use the Fourier representation of the two functional delta functions that
fix the constraints (\ref{bxdef}) and (\ref{bydef}) in Eq.~(\ref{zpartfunthree}). As a result, the path integration in  ${\cal D}\boldsymbol r_1(s)$ and ${\cal D}\boldsymbol r_2(s)$ 
reduces to a product of two path integrals in which the integrations over the conformations of the loops $C_1$ and $C_2$ are completely decoupled:
\begin{eqnarray}
  {\cal Z}[m]&=&\int
  {\cal D}\boldsymbol b^1(\boldsymbol x){\cal D}\boldsymbol b^2(\boldsymbol x)
  {\cal D}\boldsymbol c^1(\boldsymbol x){\cal D}\boldsymbol c^2(\boldsymbol x) e^{-i\sum_{i=1}^2\int d^3x\left(\boldsymbol c^i(\boldsymbol x)\cdot\boldsymbol b^i(\boldsymbol x)\right)}
\nonumber\\
&\times&\delta\left(\frac 1{4\pi}\int d^3x\int d^3y \epsilon^{\mu\nu\rho}b_\mu^1(\boldsymbol x)\frac{(x-y)_\nu}{|\boldsymbol x-\boldsymbol y|^3} b^2_\rho(\boldsymbol y)-m\right)
\delta\left(\boldsymbol\nabla\cdot \boldsymbol c^i\right)\nonumber\\
&\times&\prod_{j=1}^2\delta\left(
\boldsymbol \nabla\cdot\boldsymbol b^j(\boldsymbol x)
  \right)\prod_{i=1}^2\left[\int{\cal D}\boldsymbol r_i(s)
  e^{-\int_0^L ds\left[\frac3{2a}\boldsymbol{\dot r}_i^2(s)-i\boldsymbol{\dot r}_i(s)\cdot\boldsymbol c^i(\boldsymbol r_i(s))
      \right]}
  \right]
  \label{zpartfunfour}
\end{eqnarray}
We remark that the new fields $\boldsymbol c^1$ and $\boldsymbol c^2$ are true vector fields and not magnetic field densities. The partition function of Eq.~({\ref{zpartfunfour}}) is invariant under the gauge transformations $\boldsymbol c^i(\boldsymbol x)\longrightarrow \boldsymbol c^i(\boldsymbol x)+\boldsymbol \nabla\varphi^i(\boldsymbol x)$. As a matter of fact, the fields $\boldsymbol c^i(\boldsymbol x)$ are appearing either in the scalar product with the purely transverse fields $\boldsymbol b^i(\boldsymbol x)$ or in loops over closed paths. This invariance has been fixed using the pure Lorentz gauge.
The gauge fixing terms in this case consist of the product of functional delta functions $\prod_{i=1}^2\delta\left(\boldsymbol\nabla\cdot \boldsymbol c^i(\boldsymbol x)\right)$.
The summations over $\boldsymbol r_1(s)$ and $\boldsymbol r_2(s)$ may be tranformed into path integrals over replica fields using the formula (see \ref{appendixa} and \cite{Edwards,Kleinertbook} for its proof):
\begin{eqnarray}
  \int{\cal D}\boldsymbol r_i(s) e^{-\int_0^Lds{\cal L}(\boldsymbol {\dot r}_i(s),\boldsymbol r_i(s),\boldsymbol c^i)}=\lim_{n_i\to 0}\left[
\prod_{a=1}^{n_i}\int{\cal D}\psi_{i,a}^*{\cal D}\psi_{i,a}
\right] |\psi_{i,1}(r_{i,0},0)|^2&& \nonumber\\
  \times \exp\left\{-\sum_{a=1}^{n_i}\int d^3x\int_0^Ldt\left[ \psi_{i,a}^*\left(\frac{\partial}{\partial t}-\frac a6(\boldsymbol \nabla -i \boldsymbol c^i)^2-i\eta_i(\boldsymbol x)\right)\psi_{i,a}\right]\right\}&&\label{formbasic}
\end{eqnarray}
where
\begin{equation}
  {\cal L}(\boldsymbol {\dot r}_i(s),\boldsymbol r_i(s),\boldsymbol c^i)=\frac 3{2a}\boldsymbol{\dot r}_i^2(s)+i\boldsymbol {\dot r}_i(s) \cdot \boldsymbol c^i(\boldsymbol r_i(s))+i\eta(\boldsymbol r_i(s))\label{lagrbasic}
\end{equation}
The fields $\psi_{i,1}^*(\boldsymbol x,t),\ldots,\psi_{i,n_i}^*(\boldsymbol x,t)$ and
$\psi_{i,1}(\boldsymbol x,t),\ldots,\psi_{i,n_i}(\boldsymbol x,t)$ are $n_i-$tuples of replica complex scalar fields, while the
$\eta_i(\boldsymbol x)$'s represent real scalar fields. For the moment, we will assume that $\eta_i(\boldsymbol x)=0$.
With the help of Eq.~(\ref{formbasic}) the partition function ${\cal Z}[m]$ of Eq.~(\ref{zpartfunfour}) may be rewritten as follows:
\begin{eqnarray}
  {\cal Z}[m]&=&\lim_{n_1,n_2\to 0}\int {\cal D}(\mbox{fields})
  \delta\left(\frac 1{4\pi}\int d^3x\int d^3y \epsilon^{\mu\nu\rho}b_\mu^1(\boldsymbol x)\frac{(x-y)_\nu}{|\boldsymbol x-\boldsymbol y|^3} b^2_\rho(\boldsymbol y)-m\right) \nonumber\\
  &\times&
  \prod_{i=1}^2\left[\delta\left(\boldsymbol\nabla\cdot \boldsymbol b^i\right)
    \delta\left(\boldsymbol\nabla\cdot \boldsymbol c^i\right)
    |\psi_{i,1}(r_{i,0},0)|^2\right]e^{-i\sum_{i=1}^2\int d^3x\left(\boldsymbol c^i(\boldsymbol x)\cdot\boldsymbol b^i(\boldsymbol x)\right)}
   \nonumber\\
  &\times&  
  \exp\left\{-\sum_{i=1}^2\sum_{a_i=1}^{n_i}\int d^3x\int_0^Ldt\left[ \psi_{i,a_i}^*\left(\frac{\partial}{\partial t}-\frac a6(\boldsymbol \nabla -i \boldsymbol c^i)^2\right)\psi_{i,a_i}\right]\right\}\nonumber\\
%  &\times& \exp\left\{-\sum_{b=1}^{n_2}\int d^3x\int_0^Ldt\left[ \psi_{2,b}^*\left(\frac{\partial}{\partial t}-\frac a6(\boldsymbol \nabla -i \boldsymbol c^2)^2\right)\psi_{2,b}\right]\right\}
  \label{zpartfunfive}
\end{eqnarray}
with
\begin{eqnarray}
  \int{\cal D}(\mbox{fields})&=&\int \prod_{i=1}^2 {\cal D}\boldsymbol b^i(\boldsymbol x){\cal D}\boldsymbol c^i(\boldsymbol x)
      \prod_{a_i=1}^{n_i}{\cal D}\psi_{i,a_i}^*(\boldsymbol x,t)
           {\cal D}\psi_{i,a_i}(\boldsymbol x,t)
\end{eqnarray}
Instead of dealing with the magnetic flux densities $\boldsymbol b^i(\boldsymbol x)$,
it will be more convenient  to introduce the new vector fields $\boldsymbol a^1,\boldsymbol a^2$ performing the field transformation:
\begin{equation}
  \boldsymbol b^i(\boldsymbol x)= \boldsymbol\nabla\times \boldsymbol a^i(\boldsymbol x)\qquad\qquad i=1,2
  \label{transfone}
\end{equation}
The fields ${\boldsymbol a^i}$ are not vector field densities. The Jacobian of the transformation (\ref{transfone}) is a trivial constant. After performing the above change of variables, the functional delta function imposing the topological constraints in (\ref{zpartfunfive}) can be simplified using the identity:
\begin{equation}
  \frac 1{4\pi}\int d^3x\int d^3y \epsilon^{\mu\nu\rho}b_\mu^1(\boldsymbol x)\frac{(x-y)_\nu}{|\boldsymbol x-\boldsymbol y|^3} b^2_\rho(\boldsymbol y)=
\int d^3x\epsilon^{\mu\nu\rho}a_\mu^1(\boldsymbol x)\partial_\nu a^2_\rho(\boldsymbol x)\label{abc}
\end{equation}
To derive Eq.~(\ref{abc}) we have used the formula below:
\begin{equation}
  \frac1{4\pi}\epsilon^{\mu\nu\rho}
  \partial_\nu\epsilon_{\rho\sigma\tau}\frac{(x-y)^\tau}{|\boldsymbol x-\boldsymbol y|^3}
  =\left( \delta_\sigma^\mu-\frac{\partial_\sigma\partial^\mu}{\partial^2}\right)\delta(\boldsymbol x-\boldsymbol y)
\end{equation}
In terms of the new degrees of freedom $\boldsymbol a^i$, the partition function (\ref{zpartfunfive}) describes a model in which the interactions due to the topological constraints are propagated by a set of abelian BF fields:
\begin{eqnarray}
  {\cal Z}[m]&=&\lim_{n_1,n_2\to 0}\int {\cal D}(\mbox{fields})
  \delta\left(\frac 1{4\pi}\int d^3x\epsilon^{\mu\nu\rho}
  a_\mu^1\partial_\nu
  a^2_\rho-m\right) e^{-S_2}\nonumber\\
  &\times&  |\psi_{1,1}(r_{1,0},0)|^2|\psi_{2,1}(r_{2,0},0)|^2\prod_{i=1}^2\delta\left(\boldsymbol\nabla\cdot \boldsymbol a^i(\boldsymbol x)\right)
  \delta\left(\boldsymbol\nabla\cdot \boldsymbol c^i(\boldsymbol x)\right)
  \label{zpartfunfin}
\end{eqnarray}
with
\begin{eqnarray}
  S_2&=&i\sum_{i=1}^2\int d^3x \epsilon^{\mu\nu\rho}c^i_\mu\partial_\nu a^i_\rho\nonumber\\
  &+&
    \sum_{i=1}^2\sum_{a_i=1}^{n_i}
    \int d^3x\int_0^Ldt\left[
      \psi_{i,a_i}^*
      \frac{\partial}{\partial t}\psi_{i,a_i}+\frac a6\left|
      \left(\boldsymbol \nabla -i \boldsymbol c^i\right)\psi_{i,a_i}\right|^2\right]
\end{eqnarray}
and
\begin{eqnarray}
  \int{\cal D}(\mbox{fields})&=&\int 
      \prod_{i=1}^2{\cal D}\boldsymbol a^i(\boldsymbol x){\cal D}\boldsymbol c^i(\boldsymbol x)
      \prod_{a_i=1}^{n_i}\int{\cal D}\psi_{i,a_1}^*(\boldsymbol x,t)
           {\cal D}\psi_{i,a_i}(\boldsymbol x,t)\label{dfields}
\end{eqnarray}
This is the desired result.
As it is possible to see, the topological constraint (\ref{glndef}) imposed on the conformations of the fluctuating loops $C_1$ and $C_2$ has been replaced in the final polymer partition function by the condition:
\begin{equation}
\int d^3x \boldsymbol a^1\cdot \boldsymbol b^2=m\label{helcond}
\end{equation}
The quantity in the left hand side of Eq.~(\ref{helcond}) is called the cross-helicity of
the magnetic fluxes $\boldsymbol b^1(\boldsymbol x)=\boldsymbol\nabla\times\boldsymbol a^1(\boldsymbol x)$ and $\boldsymbol b^2 (\boldsymbol x)$.
The cross-helicity is a topological invariant widely applied in solar magnetohydrodynamics~\cite{Hornig}.

\subsection{The Laplace transformed partition function}\label{sub0}
An alternative expression of the partition function of two linked polymer rings can be derived by performing a double Laplace transform of the partition function (\ref{partfundef}) with respect to the lengths of the loops.
Taking the Laplace tranform of the partition function ${\cal Z}[m]$ of Eq.~(\ref{zpartfunfour}) with the help of formulas (\ref{app3})--(\ref{app5}) in \ref{appendixa}, we obtain the partition function $\tilde{{\cal Z}}[m]$ in the Laplace space:
\begin{eqnarray}
  {\tilde {\cal Z}}[m]&=&\lim_{n_1,n_2\to 0}\lim_{\epsilon_1,\epsilon_2\to 0}\int {\cal D}(\mbox{fields})
  \delta\left(\frac 1{4\pi}\int d^3x\epsilon^{\mu\nu\rho}
  a_\mu^1\partial_\nu
  a^2_\rho-m\right) e^{-{\tilde S}_2}\nonumber\\
  &\times&  |\psi_{1,1}(r_{1,0})|^2|\psi_{2,1}(r_{2,0})|^2\prod_{i=1}^2\delta\left(\boldsymbol\nabla\cdot \boldsymbol a^i(\boldsymbol x)\right)
  \delta\left(\boldsymbol\nabla\cdot \boldsymbol c^i(\boldsymbol x)\right)
  \label{zpartfun-LT}
\end{eqnarray}
where
\begin{eqnarray}
  {\tilde S}_2&=&i\sum_{i=1}^2\int d^3x \epsilon^{\mu\nu\rho}c^i_\mu\partial_\nu a^i_\rho\nonumber\\
  &+&
    i\sum_{i=1}^2\sum_{a_i=1}^{n_i}
    \int d^3x\psi^*_{i,a_i}\left[
      \left(z_i+i\epsilon_i\right)+\frac a6
      \left(\boldsymbol \nabla -i \boldsymbol c^i\right)^2\right]\psi_{i,a_i}\label{S2-LT}
\end{eqnarray}
and
\begin{eqnarray}
  \int{\cal D}(\mbox{fields})&=&\int
      \prod_{i=1}^2 {\cal D}\boldsymbol a^i(\boldsymbol x){\cal D}\boldsymbol c^i(\boldsymbol x)
      \prod_{a_i=1}^{n_i}\int{\cal D}\psi_{i,a_1}^*(\boldsymbol x)
           {\cal D}\psi_{i,a_i}(\boldsymbol x)\label{dfields-LT}
\end{eqnarray}

\subsection{Equivalence with the previous model of Ref.~\cite{FFIL1998}}\label{sub1}
In this Section the equivalence of the field partition function in Eq.~(\ref{zpartfunfin}) with that of the model derived in Ref.~\cite{FFIL1998} will be shown.
To this purpose we go back to Eq.~(\ref{zpartfunfive}) and express the Dirac delta function imposing the topological constraint using its Fourier representation:
\begin{eqnarray}
  {\cal Z}[m]&=&\lim_{n_1,n_2\to 0}\int_{-\infty}^{+\infty}d\lambda e^{i\lambda m}\int {\cal D}(\mbox{fields})
  e^{-i\sum_{i=1}^2\int d^3x\left(\boldsymbol c^i(\boldsymbol x)\cdot\boldsymbol b^i(\boldsymbol x)\right)}\nonumber\\
  &\times& 
  \prod_{i=1}^2\delta\left(\boldsymbol\nabla\cdot \boldsymbol b^i(\boldsymbol x)\right)
  \delta\left(\boldsymbol\nabla\cdot \boldsymbol c^i(\boldsymbol x)\right)|\psi_{i,1}(r_{i,0},0)|^2
  \nonumber\\
  &\times&\exp\left\{-\frac {i\lambda}{4\pi}\int d^3x\int d^3y \epsilon^{\mu\nu\rho}b_\mu^1(\boldsymbol x)\frac{(x-y)_\nu}{|\boldsymbol x-\boldsymbol y|^3} b^2_\rho(\boldsymbol y)\right\}\nonumber\\
  &\times&
  \exp\left\{-\sum_{a=1}^{n_1}\int d^3x\int_0^Ldt\left[ \psi_{1,a}^*\left(\frac{\partial}{\partial t}-\frac a6(\boldsymbol \nabla -i \boldsymbol c^1)^2\right)\psi_{1,a}\right]\right\}\nonumber\\
    &\times& \exp\left\{-\sum_{b=1}^{n_2}\int d^3x\int_0^Ldt\left[ \psi_{2,b}^*\left(\frac{\partial}{\partial t}-\frac a6(\boldsymbol \nabla -i \boldsymbol c^2)^2\right)\psi_{2,b}\right]\right\}
  \label{zpartfunfivebis}
\end{eqnarray}
In this new form it turns out that one of the fields $\boldsymbol b^1,\boldsymbol b^2$ may be interpreted as a Lagrange multiplier. For example, the integration over $\boldsymbol b^2$ produces in Eq.~(\ref{zpartfunfivebis}) a functional delta function enforcing the condition:
\begin{equation}
c^{2\mu}(\boldsymbol x)+\epsilon^{\sigma\tau\mu}\frac\lambda{4\pi}\int d^3y b_\sigma^1(\boldsymbol y)\frac{(x-y)_\tau}{|\boldsymbol x-\boldsymbol y|^3}=0\label{constrdeftwo}
\end{equation}
Thus, after integrating over $\boldsymbol b^2$, the partition function (\ref{zpartfunfivebis}) becomes:
\begin{eqnarray}
  {\cal Z}[m]&=&\lim_{n_1,n_2\to 0}\int_{-\infty}^{+\infty}d\lambda e^{i\lambda m}\int {\cal D}(\mbox{fields})'
  e^{-i\int d^3x\left(\boldsymbol c^1(\boldsymbol x)\cdot\boldsymbol b^1(\boldsymbol x)\right)}
    \prod_{i=1}^2|\psi_{i,1}(r_{i,0},0)|^2\nonumber\\
    &\times&\delta\left(c^{2\mu}(\boldsymbol x)+\frac {\lambda}{4\pi}\int d^3y \epsilon^{\sigma\tau\mu}b_\sigma^1(\boldsymbol y)\frac{(x-y)_\tau}{|\boldsymbol x-\boldsymbol y|^3}\right)
    \delta\left(\boldsymbol\nabla\cdot \boldsymbol b^1\right)\prod_{i=1}^2
    \delta\left(\boldsymbol\nabla\cdot \boldsymbol c^i\right)
    \nonumber\\
  &\times&
  \exp\left\{-\sum_{a=1}^{n_1}\int d^3x\int_0^Ldt\left[ \psi_{1,a}^*\left(\frac{\partial}{\partial t}-\frac a6(\boldsymbol \nabla -i \boldsymbol c^1)^2\right)\psi_{1,a}\right]\right\}\nonumber\\
    &\times& \exp\left\{-\sum_{b=1}^{n_2}\int d^3x\int_0^Ldt\left[ \psi_{2,b}^*\left(\frac{\partial}{\partial t}-\frac a6(\boldsymbol \nabla -i \boldsymbol c^2)^2\right)\psi_{2,b}\right]\right\}
  \label{zpartfunsix}
\end{eqnarray}
where, with respect to the measure ${\cal D}(\mbox{fields})$ in Eq.~(\ref{dfields}), in ${\cal D}(\mbox{fields})'$ the sum over the field $\boldsymbol b^2(\boldsymbol x)$ is missing.
At this point we perform the substitutions:
\begin{eqnarray}
  b^1_\sigma(\boldsymbol x)=\epsilon_{\sigma\rho\kappa}\partial^\rho
  B^{1\kappa}(\boldsymbol x)\qquad\qquad c^1_\mu(\boldsymbol x)=B^2_\mu(\boldsymbol x)
\end{eqnarray}
$B^2_\mu(\boldsymbol x)$ being a completely transverse vector field.
In this way, the constraint (\ref{constrdeftwo}) simplifies to:
\begin{equation}
c^2_\mu(\boldsymbol x)=\lambda B^1_\mu(\boldsymbol x)
\end{equation}
and the partition function ${\cal Z}[m]$ of Eq.~(\ref{zpartfunsix}) reduces to:
\begin{eqnarray}
  {\cal Z}[m]&=&\lim_{n_1,n_2\to 0}\int_{-\infty}^{+\infty}d\lambda e^{i\lambda m}\int {\cal D}(\mbox{fields})^{\prime\prime}
  e^{-i\epsilon^{\mu\nu\rho}\int d^3x B^2_\mu\partial_\nu B^1_\rho}
    \nonumber\\
    &\times&%\delta\left(c_\mu^2(\boldsymbol x)-\lambda \boldsymbol B^1(\boldsymbol x)\right)
    \prod_{i=1}^2\delta\left(\boldsymbol\nabla\cdot \boldsymbol B^i(\boldsymbol x)\right)|\psi_{1,1}(r_{1,0},0)|^2|\psi_{2,1}(r_{2,0},0)|^2
\nonumber\\
  &\times&
  \exp\left\{-\sum_{a=1}^{n_1}\int d^3x\int_0^Ldt\left[ \psi_{1,a}^*\left(\frac{\partial}{\partial t}-\frac a6(\boldsymbol \nabla -i \lambda\boldsymbol B^2)^2\right)\psi_{1,a}\right]\right\}\nonumber\\
    &\times& \exp\left\{-\sum_{b=1}^{n_2}\int d^3x\int_0^Ldt\left[ \psi_{2,b}^*\left(\frac{\partial}{\partial t}-\frac a6(\boldsymbol \nabla -i\lambda \boldsymbol B^1)^2\right)\psi_{2,b}\right]\right\}
  \label{zpartfunseven}
\end{eqnarray}
with
\begin{equation}
  {\cal D}(\mbox{fields})^{\prime\prime}={\cal D}\boldsymbol B^1(\boldsymbol x){\cal D}\boldsymbol B^2(\boldsymbol x)
  \prod_{i=1}^2\prod_{a_i=1}^{n_i}{\cal D}\psi_{i,a_i}^*(\boldsymbol x,t){\cal D}\psi_{i,a_i}(\boldsymbol x,t)
\end{equation}
The above partition function coincides with that of the model derived in \cite{FFIL1998}.
\subsection{Statistical mechanical interpretation of the derivation of Eq.~(\ref{zpartfunseven})}\label{sub2}
First, we note that the partition function (\ref{zpartfunfin}) can be regarded as a statistical sum in the microcanonical ensemble where the energy has been replaced by the cross-helicity. Indeed, we may rewrite Eq.~(\ref{zpartfunfin}) in schematic form as follows:
\begin{equation}
  {\cal Z}[m]=\int{\cal D} X{\cal D}Y\delta\left(H(X)-m\right)\exp\left\{
-iX\cdot Y+S(Y)
  \right\}
\end{equation}
where $X$ represents a generic configuration of the fields $\boldsymbol a^1(\boldsymbol x)$ and $\boldsymbol a^2(\boldsymbol x)$, $Y$ denotes the configurations of the remaining fields and $H(X)$ is the cross-helicity. Moreover $X\cdot Y=\sum_{i=1}^2\epsilon^{\mu\nu\rho}\int d^3x a_\mu^i\partial_\nu c^i_\rho$ and $S(Y)$ contains all the contributions of the fields $\boldsymbol c^i(\boldsymbol x)$, $\psi^*_{i,a_i}(\boldsymbol x,t)$
and $\psi_{i,a_i}(\boldsymbol x,t)$.
In order to pass to the canonical partition function, we introduce the  fictitious thermodynamic temperature $\lambda^{-1}$ and sum ${\cal Z}[m]$ over all possible values of the {\em energy} levels $m$:
\begin{equation}
{\cal Z}[\lambda]=\int_{-\infty}^{+\infty}e^{-i\frac m\lambda}{\cal Z}[m]\label{zpartfuneight}
\end{equation}
The results of the integration over $m$ is:
\begin{equation}
  {\cal Z}[\lambda]=\int{\cal D}X{\cal D}Ye^{-i\frac{H(X)}{\lambda}}\exp\left\{-iX\cdot Y+S(Y)
  \right\}\label{zcanonical}
\end{equation}
${\cal Z}[\lambda]$ can be considered as the canonical partition function. The only difference in the passage from the microcanonical ensemble to the canonical one is dictated by the fact that in the present case the energy is replaced by the cross-helicity. In statistical mechanics the energy is supposed to be bounded from below. The cross-helicity instead can take arbitrary negative values.  For this reason in Eq.~(\ref{zpartfuneight}) ${\cal Z}[\lambda]$ is obtained from ${\cal Z}[m]$ via a Fourier transformation and not via a Laplace transform.

In Eq.~(\ref{zcanonical}) the integration over the $X$ degrees of freedom, i.~e. the fields $\boldsymbol a^1(\boldsymbol x)$ and $\boldsymbol a^2(\boldsymbol x)$, becomes straightforward. The result of this integration is the partition function of Eq.~(\ref{zpartfunseven}). It is thus possible to say that the model of linked two polymers obtained here, see Eq.~(\ref{zpartfunfin}) and that of Ref.~\cite{FFIL1998} are the same partition function represented in two different statistical ensembles.
\subsection{Solving the constraint $\frac 1{4\pi}\int d^3 x \epsilon^{\mu\nu\rho}a^1_\mu\partial_\nu a^2_\rho = m$ }\label{sub3}
As we have seen, the field theory version of the constraint (\ref{glndef}) is a condition on the cross-helicity of the fields $\boldsymbol a^i(\boldsymbol x)$, $i=1,2$:
\begin{equation}
  \frac 1{4\pi}\int d^3 x \epsilon^{\mu\nu\rho}a^1_\mu\partial_\nu a^2_\rho = m\label{topconstrFTversion}
\end{equation}
In order to find the solution of the above equation, we consider two non-fluctuating (i. e. static) loops ${\bar C}_1$ and ${\bar C}_2$ that are concatenated in such a way that the link formed by them has Gauss linking number equal to $m$:
\begin{equation}
\chi({\bar C}_1,{\bar C}_2)=\oint_{{\bar C}_1} \frac{dx_1^\mu}{4\pi}\oint_{{\bar C}_2} dx^\nu_2\frac{(x_2-x_1)^\rho}{|\boldsymbol x_2-\boldsymbol x_1 |}=m
\end{equation}
Next, we rewrite the above equation in a field theory form:
\begin{equation}
\int \frac{d^3x}{4\pi} \epsilon^{\mu\nu\rho}A^1_\mu\partial_\nu A^2_\rho=m
\end{equation}
Indeed, choosing the configurations of the classical fields $\boldsymbol A^1(\boldsymbol x),
\boldsymbol A^2(\boldsymbol x)$ as follows:
\begin{equation}
  A^{1\,\mu}(\boldsymbol x)=\oint_{{\bar C}_1}dx_1^\mu\delta(\boldsymbol x-\boldsymbol x_1)\qquad\mbox{and}\qquad
   A^{2\,\mu}(\boldsymbol x)=\oint_{{\bar C}_2}\frac{dx_2^\mu}{|\boldsymbol x-\boldsymbol x_2|}\label{Asolns}
\end{equation}
it is easy to show that:
\begin{equation}
\oint_{{\bar C}_1} \frac{dx_1^\mu}{4\pi}\oint_{{\bar C}_2} dx^\nu_2\frac{(x_2-x_1)^\rho}{|\boldsymbol x_2-\boldsymbol x_1 |} = \int \frac{d^3x}{4\pi} \epsilon^{\mu\nu\rho}A^1_\mu\partial_\nu A^2_\rho
\end{equation}
Eq.~(\ref{Asolns}) provides the solutions  $\boldsymbol A^1(\boldsymbol x),
\boldsymbol A^2(\boldsymbol x)$ of the field theoretical version of the topological constraint given in Eq.~(\ref{topconstrFTversion}).

Performing in the partition function (\ref{zpartfunfin}) the change of variables:
\begin{equation}
\boldsymbol a^i(\boldsymbol x)=\boldsymbol A^i(\boldsymbol x)+\delta\boldsymbol a^i(\boldsymbol x)
\end{equation}
where $\delta\boldsymbol a^i(\boldsymbol x)$ is a small perturbation of the static field configurations
$\boldsymbol A^i(\boldsymbol x)$,
it is possible to study the fluctuations around the fixed conformations ${\bar C}_1$ and ${\bar C}_2$ of two polymer rings forming a given link with Gauss linking number equal to $m$. 
\section{The case of a higher order link invariant }\label{sec3}
In this Section the previous strategy to build field theory models of linked polymer rings will be generalized to higher order numerical topological invariants. In particular, it will be considered the following invariant which takes into account of the topological relations between three rings $C_1,C_2,C_3$:
\begin{equation}
\tau_3(C_1,C_2,C_3)=\int d^3x\epsilon_{ijk}\epsilon^{\mu\nu\rho}{\cal B}^i_\mu(\boldsymbol x){\cal B}^j_\nu(\boldsymbol y) {\cal B}^k_\rho(\boldsymbol z)\label{massey}
\end{equation}
where
\begin{equation}
  {\cal B}^i_\mu(\boldsymbol x)=\frac{\epsilon^\mu_{\sigma\tau}}{4\pi}\oint_{C_i}dr_i^\sigma\frac {(x-r_i)^\tau}{|\boldsymbol x-\boldsymbol r_i|^3}
\end{equation}
In the above equation the indices $i,j,k=1,2,3$ number the loops $C_1,C_2,C_3$.
$\tau_3(C_1,C_2,C_3)$ is a special case of a class of invariants due to Massey. It has already been proposed to distinguish the topological states of the solar magnetic field \cite{Hornig} and of a system of linked polymers \cite{nova,non-semisimple}. It has been shown in \cite{non-semisimple} that the expression of $\tau_3(C_1,C_2,C_3)$ may be isolated from the amplitudes of a Chern-Simons field theory with non semi-simple group of gauge symmetry.
As in the case of the Gauss linking number, we rewrite 
$\tau_3(C_1,C_2,C_3)$  using magnetic flux densities $\boldsymbol b^1,\boldsymbol b^2$ and $\boldsymbol b^3$:
\begin{equation}
  \tau_3(\boldsymbol b^1,\boldsymbol b^2,\boldsymbol b^3)=\epsilon_{ijk}\int d^3 x d^3y d^3z b^i_{\kappa_1}(\boldsymbol x)
  b^j_{\eta_1}(\boldsymbol y)  b^k_{\lambda_1}(\boldsymbol z) I^{\kappa_1\eta_1\lambda_1}(\boldsymbol x,\boldsymbol y,\boldsymbol z)\label{tau3cinb}
\end{equation}
with
\begin{equation}
  I^{\kappa_1\eta_1\lambda_1}(\boldsymbol x,\boldsymbol y,\boldsymbol z)=\int d^3\omega\epsilon_{\mu\nu\rho}\epsilon^{\mu\kappa_1\kappa_2} \epsilon^{\nu\eta_1\eta_2}\epsilon^{\rho\lambda_1\lambda_2}
  \frac{(\omega -x)_{\kappa_2}}{|\boldsymbol \omega-\boldsymbol x|^3}
  \frac{(\omega -y)_{\eta_2}}{|\boldsymbol \omega-\boldsymbol y|^3}
    \frac{(\omega -z)_{\lambda_2}}{|\boldsymbol \omega-\boldsymbol z|^3}
  \end{equation}
It is easy to check that the relation
\begin{equation}
\tau_3(C_1,C_2,C_3)=\tau_3(\boldsymbol b^1,\boldsymbol b^2,\boldsymbol b^3)
\end{equation}
is valid provided:
\begin{equation}
\boldsymbol b^i(\boldsymbol x)=\int_0^Lds\boldsymbol {\dot r}_i(s)\delta(\boldsymbol x-\boldsymbol r_i(s))\label{bi3}
\end{equation}
In the following we will consider the amplitude of an observable $\cal O$ of the general form:
\begin{equation}
  {\cal O}=\sum_{i,j,k=1}^3\int_0^Lds_i\int_0^Lds_j'\oint_0^Lds_k^{\prime\prime}f(\boldsymbol r_i(s_i),
  \boldsymbol r_j(s_j'),\boldsymbol r_k
  (s_k^{\prime\prime}))A^{ijk}\label{obsgenform}
\end{equation}
The $A^{ijk}$'s, $i=1,2,3$, are constant coefficients.
%For example, the square distance between the centers of mass of the loops $C_1$ and $C_2$:
%\begin{equation}
%  \left|
%\int_0^Lds_1\boldsymbol r_1(s_1)-\int_0^Lds_2\boldsymbol r_2(s_2)
%  \right|^2={\cal O}_{11}+{\cal O}_{22}-2{\cal O}_{12}\label{hhh}
%\end{equation}
%with:
%\begin{equation}
%{\cal O}_{ij}=\int_0^Lds_i\int_0^Lds_j'\boldsymbol r_i(s_i)\cdot\boldsymbol r_j(s'_j)
%\end{equation}
%may be expressed in the form of Eq.~(\ref{obsgenform})
%with the special choice $f(\boldsymbol r_1,\boldsymbol r_2,\boldsymbol r_3)=\sum_{i,j,k=1}^3A_{i,j,k}\boldsymbol %r_i\cdot\boldsymbol r_j$ and $A_{ijk}=\delta{ij}\delta_{k1}$.
 Similarly to what has been done in the case of the Gauss linking number, we eliminate the contour integrations over $C_1,C_2,C_3$ in the observable introducing fields. In the present case it is possible to rewrite the expression of $\cal O$ using a set of scalar fields $\phi_1,\phi_2,\phi_3$ such that:
\begin{equation}
\phi_i(\boldsymbol x)=\int_0^Lds_i\delta(\boldsymbol x -\boldsymbol r_i(s_i))
\end{equation}
The form of $\cal O$ in terms of the fields $\phi_i'$s is:
\begin{equation}
{\cal O}=\sum_{i,j,k=1}^3\int d^3xd^3yd^3zf(\boldsymbol x,\boldsymbol y,\boldsymbol z)\phi_i(\boldsymbol x)\phi_j(\boldsymbol y)\phi_k(\boldsymbol z)A^{ijk}
\end{equation}
We are now ready to consider the expectation value $\left\langle
{\cal O}
\right\rangle$ for a system of three polymer rings whose fluctuations are constrained by the condition $\tau_3(C_1,C_2,C_3)=m$:
\begin{eqnarray}
  \left\langle
{\cal O}
\right\rangle&=&\prod_{i=1}^3\left[
\int {\cal D}\boldsymbol r_i(s){\cal D}\boldsymbol b^i(\boldsymbol x){\cal D}\phi_i(\boldsymbol x)
\right]\delta\left(\tau_3(\boldsymbol b^1,\boldsymbol b^2,\boldsymbol b^3)-m\right)
\nonumber\\
&\times&\left(
\sum_{i,j,k=1}^3\int d^3xd^3yd^3z f(\boldsymbol x,\boldsymbol y,\boldsymbol z)\phi_i(\boldsymbol x)\phi_j(\boldsymbol y)\phi_k(\boldsymbol z)A^{ijk}
\right)
\nonumber\\
&\times&\prod_{i=1}^3\left[
  \delta\left(\phi_i(\boldsymbol x)-\int_0^Lds\delta(\boldsymbol x-\boldsymbol r_i(s))\right)\right]
\prod_{i=1}^3\left[\delta\left(
\boldsymbol \nabla\cdot\boldsymbol b^i(\boldsymbol x)
  \right)
  \right]
\nonumber\\
  &\times& \prod_{i=1}^3\left[ \delta\left(
  \boldsymbol b_i(\boldsymbol x)-\int_0^Lds\boldsymbol {\dot r}_i(s)\delta(\boldsymbol x-\boldsymbol r_i(s))
  \right)
  \right]e^{-S_{pol}^3}\label{Oone}
\end{eqnarray}
The definition of $S_{pol}^3$ has been provided in Eq.~(\ref{spol}).
To pass to the Fourier representation of the functional Dirac delta functions in the last two lines of Eq.~(\ref{Oone}), we introduce the new fields $\eta_i(\boldsymbol x)$ and $\boldsymbol c^i(\boldsymbol x)$.
They are respectively the Fourier conjugate fields of $\phi_i(\boldsymbol x)$ and $\boldsymbol b^i(\boldsymbol x)$.
As an upshot, it is possible to rewrite Eq.~(\ref{Oone}) as follows:
\begin{eqnarray}
  \left\langle
{\cal O}
\right\rangle&=&\prod_{i=1}^3\left[
  \int {\cal D}\boldsymbol b^i(\boldsymbol x)\boldsymbol c^i(\boldsymbol x)
       {\cal D}\phi_i(\boldsymbol x){\cal D}\eta_i(\boldsymbol x)
\right]\delta\left(\tau_3(\boldsymbol b^1,\boldsymbol b^2,\boldsymbol b^3)-m\right)
%e^{-\sum_{i=1}^3\int_0^L\frac{3}{2a}\boldsymbol{\dot r}_i^2(s)}
\nonumber\\
&\times&\left(
\sum_{i,j,k=1}^3\int d^3xd^3yd^3z f(\boldsymbol x,\boldsymbol y,\boldsymbol z)\phi_i(\boldsymbol x)\phi_j(\boldsymbol y)\phi_k(\boldsymbol z)A^{ijk}
\right)\nonumber\\
&\times&
\left[\prod_{i=1}^3\delta\left(
\boldsymbol \nabla\cdot\boldsymbol b^i
\right)\right]
\left[\prod_{i=1}^3\delta\left(
\boldsymbol \nabla\cdot\boldsymbol c^i
\right)\right ]\nonumber\\
&\times&
\exp\left\{-i\sum_{i=1}^3\int d^3x\left(\phi_i(\boldsymbol x)\eta_i(\boldsymbol x)+\boldsymbol c^i(\boldsymbol x)\cdot\boldsymbol b^i(\boldsymbol x)\right)\right\}      {\cal Z}\left[\eta_i,\boldsymbol c^i\right]
\label{Otwo}
  \end{eqnarray}
where
\begin{equation}
  {\cal Z}\left[
\eta_i,\boldsymbol c^i
    \right]=\prod_{i=1}^3\int{\cal D}\boldsymbol r_i(s)
\exp\left\{
-\int_0^Lds\left[
\frac 3{2a}\boldsymbol{\dot r}_i^2(s)-i\boldsymbol {\dot r_i}(s)\cdot\boldsymbol c^i(\boldsymbol r_i(s))-i\eta_i(\boldsymbol r_i(s))
  \right]
\right\}\label{Otwoaux}
\end{equation}
The integrations over the polymer conformations $\boldsymbol r_i(s)$ can be performed using Eq.~(\ref{formbasic}).
The result is that $\langle\cal O\rangle$ is mapped into the amplitude of a local field theory:
\begin{eqnarray}
  \left\langle
{\cal O}
\right\rangle&=&\lim_{n_1,n_2,n_3\to 0}
\prod_{i=1}^3\left[
  \int {\cal D}\boldsymbol b^i(\boldsymbol x)\boldsymbol c^i(\boldsymbol x)
       {\cal D}\phi_i(\boldsymbol x){\cal D}\eta_i(\boldsymbol x)\prod_{a_i=1}^{n_i}\left[
         {\cal D}\psi_{i,a_i}^{*}(\boldsymbol x,t)
         {\cal D}\psi_{i,a_i}(\boldsymbol x,t)
       \right]\right]
\nonumber\\
&\times&\left(
\sum_{i,j,k=1}^3\int d^3xd^3yd^3z f(\boldsymbol x,\boldsymbol y,\boldsymbol z)\phi_i(\boldsymbol x)\phi_j(\boldsymbol y)\phi_k(\boldsymbol z)A^{ijk}
\right)\delta\left(\tau_3(\boldsymbol b^1,\boldsymbol b^2,\boldsymbol b^3)-m\right)\nonumber\\
&\times&\exp\left\{-i\sum_{i=1}^3\int d^3x\left(\phi_i(\boldsymbol x)\eta_i(\boldsymbol x)+\boldsymbol c^i(\boldsymbol x)\cdot\boldsymbol b^i(\boldsymbol x)\right)\right\}
%\right)\right]
\nonumber\\
&\times&\exp\left\{
-\sum_{i=1}^3\sum_{a_i=1}^{n_i}\int d^3x\int_0^Ldt\psi_{i,a_i}^{*}\left[
  \frac\partial{\partial t}-\frac a6\left(
\boldsymbol\nabla -i \boldsymbol c^i
  \right)^2-i\eta_i(\boldsymbol x)
  \right]\psi_{i,a_i}
\right\}
\nonumber\\
&\times&\prod_{i=1}^3|\psi_{i,1}(\boldsymbol r_{i,0},0)|^2\delta\left(
\boldsymbol \nabla\cdot\boldsymbol b^i(\boldsymbol x)\right)\delta\left(
\boldsymbol \nabla\cdot\boldsymbol c^i(\boldsymbol x)\right)
\label{Othree}
  \end{eqnarray}
The fields $\eta_i(\boldsymbol x)$ are Lagrange multipliers. The integration over these degree of freedom results in the appearence in the right hand side of Eq.~(\ref{Othree}) of a functional Dirac delta functions imposing the conditions:
\begin{equation}
\phi_i(\boldsymbol x)=\sum_{a_i=1}^{n_i}\int_0^Ldt |\psi_{i,a_i}(\boldsymbol x,t)|^2\qquad\qquad i=1,2,3
\end{equation}
As a consequence, after the summation over the fields $\phi_i(\boldsymbol x)$ and $\eta_i(\boldsymbol x)$ we obtain as a result:
\begin{eqnarray}
  \left\langle
{\cal O}
\right\rangle&=&\lim_{n_1,n_2,n_3\to 0}
\prod_{i=1}^3\left[
  \int {\cal D}\boldsymbol b^i(\boldsymbol x)\boldsymbol c^i(\boldsymbol x)
       \prod_{a_i=1}^{n_i}\left[
         {\cal D}\psi_{i,a_i}^{*}(\boldsymbol x,t)
         {\cal D}\psi_{i,a_i}(\boldsymbol x,t)
       \right]\right]
\nonumber\\
&\times&\left(
\sum_{i,j,k=1}^3A^{ijk}\int d^3xd^3yd^3z f(\boldsymbol x,\boldsymbol y,\boldsymbol z)\sum_{a_i=1}^{n_i}\int_0^Ldt|\psi_{i,a_i}(\boldsymbol x,t)|^2\right.\nonumber\\
&\times&\left.
\sum_{a_j=1}^{n_j}\int_0^Ldt'|\psi_{j,a_j}(\boldsymbol y,t')|^2
\sum_{a_k=1}^{n_k}\int_0^Ldt^{\prime\prime}|\psi_{k,a_k}(\boldsymbol z,t^{\prime\prime})|^2
\right)
\nonumber\\
&\times&\delta\left(\tau_3(\boldsymbol b^1,\boldsymbol b^2,\boldsymbol b^3)-m\right) \prod_{i=1}^3|\psi_{i,1}(\boldsymbol r_{i,0},0)|^2\left[\prod_{i=1}^3
\delta\left(
\boldsymbol \nabla\cdot\boldsymbol b^i
\right)
  \delta\left(
\boldsymbol \nabla\cdot\boldsymbol c^i
\right)\right]\nonumber
\\
&\times&
\exp\left\{-i\sum_{i=1}^3\int d^3x\boldsymbol c^i(\boldsymbol x)\cdot\boldsymbol b^i(\boldsymbol x)\right\}
\nonumber\\
&\times&\exp\left\{
-\sum_{i=1}^3\sum_{a_i=1}^{n_i}\int d^3x\int_0^Ldt\psi_{i,a_i}^{*}\left[
  \frac\partial{\partial t}-\frac a6\left(
\boldsymbol\nabla -i \boldsymbol c^i
  \right)^2
  \right]\psi_{i,a_i}
\right\}
\label{Ofour}
  \end{eqnarray}
Finally, we perform the field transformations:
\begin{equation}
\boldsymbol b^i(\boldsymbol x)=\boldsymbol \nabla\times\boldsymbol a^i(\boldsymbol x)\qquad\qquad i=1,2,3
\end{equation}
The Jacobian of this transformation is a trivial constant. We require also that the fields $\boldsymbol  a^i(\boldsymbol x)$ are quantized in the pure Lorentz gauge in which:
\begin{equation}
\boldsymbol \nabla\cdot\boldsymbol a^i(\boldsymbol x)=0\qquad\qquad i=1,2,3
\end{equation}
After a straightforward calculation it is possible to check that in terms of the new fields $\boldsymbol  a^i(\boldsymbol x)$ the expectation value of the observable $\cal O$
takes the following form:
\begin{eqnarray}
  \left\langle
{\cal O}
\right\rangle&=&\lim_{n_1,n_2,n_3\to 0}
\prod_{i=1}^3\left[
  \int {\cal D}\boldsymbol  a^i(\boldsymbol x)\boldsymbol c^i(\boldsymbol x)
       \prod_{a_i=1}^{n_i}\left[
         {\cal D}\psi_{i,a_i}^{*}(\boldsymbol x,t)
         {\cal D}\psi_{i,a_i}(\boldsymbol x,t)
       \right]\right]e^{-S_3}
\nonumber\\
&\times&\delta\left(\tau_3(\boldsymbol a^1,\boldsymbol  a^2,\boldsymbol  a^3)-m\right) \left[\prod_{i=1}^3\delta\left(
\boldsymbol \nabla\cdot\boldsymbol c^i(\boldsymbol x)
\right)\delta\left( \boldsymbol\nabla\cdot \boldsymbol a(\boldsymbol x)\right)\right]\nonumber\\
&\times&\left(
\sum_{i,j,k=1}^3A^{ijk}\int d^3xd^3yd^3z f(\boldsymbol x,\boldsymbol y,\boldsymbol z)\left(\sum_{a_i=1}^{n_i}\int_0^Ldt|\psi_{i,a_i}(\boldsymbol x,t)|^2\right)\right.\nonumber\\
&\times&\left.
\left(\sum_{a_j=1}^{n_j}\int_0^Ldt'|\psi_{j,a_j}(\boldsymbol y,t')|^2\right)
\left(\sum_{a_k=1}^{n_k}\int_0^Ldt^{\prime\prime}|\psi_{k,a_k}(\boldsymbol z,t^{\prime\prime})|^2\right)
\right)
\nonumber\\
&\times&\prod_{i=1}^3|\psi_{i,1}(\boldsymbol r_{i,0},0)|^2
\label{Ofinal}
  \end{eqnarray}
where
\begin{equation}
S_3=i\sum_{i=1}^3\int d^3x\epsilon^{\mu\nu\rho} c^i_\mu\partial_\nu a^i_\rho
+\sum_{i=1}^3\sum_{a_i=1}^{n_i}\int d^3x\int_0^Ldt\psi_{i,a_i}^{*}\left[
  \frac\partial{\partial t}-\frac a6\left(
\boldsymbol\nabla -i \boldsymbol c^i
  \right)^2
  \right]\psi_{i,a_i}
\end{equation}
and
\begin{equation}
\tau_3(\boldsymbol a^1,\boldsymbol  a^2,\boldsymbol  a^3)=\int d^3\omega\epsilon_{ijk}\epsilon^{\mu\nu\rho} a^i_\mu(\boldsymbol \omega) a^j_\nu(\boldsymbol \omega) a^k_\rho(\boldsymbol \omega)\label{thirdorderinv}
\end{equation}
\section{The moments of the Gauss linking number as amplitudes of a free field theory}\label{sec4}
The starting point of this Section is the partition function ${\tilde {\cal Z}}[m]$ of Eq.~(\ref{zpartfun-LT}).
In the action ${\tilde S}_2$ of Eq.~(\ref{S2-LT}) we  isolate the contributions that are linear and quadratic in $\boldsymbol c^i$:
\begin{eqnarray}
  {\tilde S}_2&=&i\sum_{i=1}^2\int d^3x 
  c^i_\mu\Bigg\{
  \epsilon^{\mu\nu\rho}
  \partial_\nu a^i_\rho
  -\sum_{a_i=1}^{n_i}\frac
  a6\left[i\left(\psi^*_{i,a_i}\partial^\mu\psi_{i,a_i}-\partial^\mu\psi^*_{i,a_i}\psi_{i,a_i}\right)
  \right.\nonumber\\
  &+&\phantom{\sum_{a_i=1}^{n_i}} c^{i,\mu}|\psi_{i,a_i}^2|\Bigg\}+ i \sum_{i=1}^2\sum_{a_i=1}^{n_i}
  \left[
(z_i+i\epsilon_i)|\psi_{i,a_i}|^2-\frac a6|\boldsymbol \nabla\psi_{i,a_i}|^2
    \right]\label{actionS2ci}
\end{eqnarray}
At this point we put
\begin{equation}
J_i^\mu =-\sum_{a_i=1}^{n_i}\frac
a6\left[i\left(\psi^*_{i,a_i}\partial^\mu\psi_{i,a_i}-\partial^\mu\psi^*_{i,a_i}\psi_{i,a_i}\right)+c^{i,\mu}|\psi_{i,a_i}|^2 \right]_\perp\label{currentJ}
\end{equation}
so that Eq.~(\ref{actionS2ci}) becomes:
\begin{eqnarray}
  {\tilde S}_2&=&i\sum_{i=1}^2\int d^3x 
  c^i_\mu\big\{
  \epsilon^{\mu\nu\rho}
  \partial_\nu a^i_\rho +J_i^\mu
  \big\}\nonumber\\
  &+& i \sum_{i=1}^2\sum_{a_i=1}^{n_i}
  \left[
(z_i+i\epsilon_i)|\psi_{i,a_i}|^2-\frac a6|\boldsymbol \nabla\psi_{i,a_i}|^2
    \right]\label{S2curr}
\end{eqnarray}
We note that the $\boldsymbol c^i$ fields are completely transverse vector fields, so that only the transverse part of $\boldsymbol J_i$ is selected when computing $\int d^3 x c^i_\mu J^\mu_i$. For that reason in Eq.~(\ref{currentJ}) only the purely transverse components of the currents $\boldsymbol J_i$ are to be chosen, as it is stressed by the presence of the subscript $\perp$ in the right hand side of Eq.~(\ref{currentJ}).
The solutions $a^{i(cl)}_\rho(\boldsymbol x)$ of the equations:
\begin{equation}
\epsilon^{\mu\nu\rho}
  \partial_\nu a^i_\rho=-J_i^\mu\qquad\qquad i=1,2
\end{equation}
are:
\begin{equation}
a^{i(cl)}_\rho(\boldsymbol x)=\frac1{4\pi}\epsilon_{\rho\sigma\tau}\int d^3y\frac{(x-y)^\tau}{|\boldsymbol x-\boldsymbol y|^3}J_i^\sigma(\boldsymbol x)\label{airhosol1}
\end{equation}
It is easy to show that in Eq.~(\ref{airhosol1}) only the purely transverse components of the currents $J_i^\sigma(\boldsymbol x)$ are present. Indeed, for any longitudinal vector field $\partial^\sigma\phi(\boldsymbol y)$ the following identity holds:
\begin{equation}
  \int d^3y\epsilon_{\rho\sigma\tau}\frac{(x-y)^\tau}{|\boldsymbol x-\boldsymbol y|^3}\partial^\sigma\phi(\boldsymbol y)=0
\end{equation}
As a consequence, from now on the subscript $\perp$ in the expression of the currents $J_i^\mu(\boldsymbol x)$ of Eq.~(\ref{currentJ}) will be dropped.
Performing in Eq.~(\ref{S2curr}) the shift:
\begin{equation}
a_\rho^i(\boldsymbol x)=a_\rho^{i\prime}(\boldsymbol x)- a_\rho^{i(cl)}(\boldsymbol x)
  \label{atransf}
\end{equation}
we may rewrite the partition function ${\tilde{\cal Z}}[m]$ of Eq.~(\ref{zpartfun-LT}) as follows:
\begin{eqnarray}
  {\tilde {\cal Z}}[m]&=&\lim_{n_1,n_2\to 0}\lim_{\epsilon_1,\epsilon_2\to 0}\int\prod_{i=1}^2\left[{\cal D}
  \boldsymbol a^{i\prime}(\boldsymbol x) {\cal D}\boldsymbol c^i(\boldsymbol x) 
      \prod_{a_i=1}^{n_i}\left(\int{\cal D}\psi_{i,a_1}^*(\boldsymbol x)
           {\cal D}\psi_{i,a_i}(\boldsymbol x)\right)\right]\nonumber\\
&\times&  \delta\left(\frac 1{4\pi}\int d^3x\epsilon^{\mu\nu\rho}
  \left(a_\mu^{1\prime}-a_\mu^{1(cl)}\right)\partial_\nu\left(
  a^{2\prime}_\rho-a_\rho^{2(cl)}\right)-m\right) e^{-{\tilde S}_2^\prime}\nonumber\\
  &\times&  
  \prod_{i=1}^2|\psi_{i,1}(r_{i,0})|^2|\delta\left(\boldsymbol\nabla\cdot \boldsymbol a^{i\prime}(\boldsymbol x)\right)
  \delta\left(\boldsymbol\nabla\cdot \boldsymbol c^i(\boldsymbol x)\right)
  \label{zpartfun-LT3}
\end{eqnarray}
where
\begin{eqnarray}
  {\tilde S}_2^\prime&=&i\sum_{i=1}^2\int d^3x \epsilon^{\mu\nu\rho}c^i_\mu\partial_\nu a^{i\prime}_\rho\nonumber\\
  &+&
    i\sum_{i=1}^2\sum_{a_i=1}^{n_i}
    \int d^3x\left[
      \left(z_i+i\epsilon_i\right)|\psi_{i,a_i}|^2-\frac a6
      |\boldsymbol \nabla \psi_{i,a_i}|^2\right]\label{S2P-curr}
\end{eqnarray}
In order to obtain Eqs.~(\ref{zpartfun-LT3}) and (\ref{S2P-curr}) we have used the fact that the integration
measure  of the  $\boldsymbol a^i$ fields is left invariant by the field transformations (\ref{atransf}). Moreover, since $\boldsymbol \nabla\cdot \boldsymbol a^{i(cl)}=0$, the transversality condition of the fields
 $\boldsymbol a^i(\boldsymbol x)$
imposed with the help of Dirac delta functions in Eq.~(\ref{zpartfun-LT}), applies also to the fields
 $\boldsymbol a^{i\prime}(\boldsymbol x)$.

In the action (\ref{S2P-curr}) the $\boldsymbol c^i(\boldsymbol x)$ fields are Lagrange multipliers enforcing the condition:
\begin{equation}
\epsilon^{\mu\nu\rho}\partial_\nu a^{i\prime}_\rho(\boldsymbol x)=0
\end{equation}
Together with the transversality requirement $\boldsymbol \nabla \cdot\boldsymbol a^{i\prime}(\boldsymbol x)=0$, this implies that the fields $\boldsymbol a^{i\prime}(\boldsymbol x)$ may be easily integrated out in the partition function of Eq.~(\ref{zpartfun-LT3}). As a result of this integration the fields $\boldsymbol a^{i\prime}(\boldsymbol x)$ can be set everywhere to be equal to zero. The final expression of the partition function ${\tilde {\cal Z}}[m]$ becomes:
\begin{eqnarray}
  {\tilde {\cal Z}}[m]&=&\lim_{n_1,n_2\to 0}\lim_{\epsilon_1,\epsilon_2\to 0}\int
        \prod_{i=1}^2\prod_{a_i=1}^{n_i}\left(\int{\cal D}\psi_{i,a_1}^*(\boldsymbol x)
           {\cal D}\psi_{i,a_i}(\boldsymbol x)\right)\nonumber\\
&\times&  \delta\left(\frac 1{4\pi}\int d^3x\epsilon^{\mu\nu\rho}
           a_\mu^{1(cl)}\partial_\nu a_\rho^{2(cl)}-m\right)
           \prod_{i=1}^2|\psi_{i,1}(r_{i,0})|^2|
           e^{-{\tilde S}_2^{\prime\prime}}  \label{zpartfun-LTfin}
\end{eqnarray}
with
\begin{eqnarray}
  {\tilde S}_2^{\prime\prime}&=&
    i\sum_{i=1}^2\sum_{a_i=1}^{n_i}
    \int d^3x\left[
      \left(z_i+i\epsilon_i\right)|\psi_{i,a_i}|^2-\frac a6
      |\boldsymbol \nabla \psi_{i,a_i}|^2\right]\label{S2Pcurrfin}
\end{eqnarray}
Note that, while the action $S_2^{\prime\prime}$ contains only the fields $\psi_{i,a_i}^*,\psi_{i,a_i}$ and there are only quadratic terms with no coupling,  Eq.~(\ref{zpartfun-LTfin}) does not represent the partition function of a free model. As a matter of fact, the fields $\psi_{i,a_i}^*,\psi_{i,a_i}$ interact via the Dirac delta function that enforce the constraint:
\begin{equation}
  \frac 1{4\pi}\int d^3x\epsilon^{\mu\nu\rho}
  a_\mu^{1(cl)}\partial_\nu a_\rho^{2(cl)}=m
\end{equation}
This interaction becomes evident after passing to the Fourier representation of the delta function:
\begin{equation}
\delta\left(\frac 1{4\pi}\int d^3x\epsilon^{\mu\nu\rho}
a_\mu^{1(cl)}\partial_\nu a_\rho^{2(cl)}-m\right)=\int_{-\infty}^{+\infty} d\lambda
e^{-i\lambda\left[\frac 1{4\pi}\int d^3x\epsilon^{\mu\nu\rho}
  a_\mu^{1(cl)}\partial_\nu a_\rho^{2(cl)}-m\right]}
\end{equation}
Looking at the expression of the fields $a_\mu^{i(cl)}$ given by Eqs.~(\ref{currentJ}) and (\ref{airhosol1}), it is easy to realize that the exponent in the right hand side of the above equation contains quartic interactions of the fields  $\psi_{i,a_i}^*,\psi_{i,a_i}$.

Even if the partition function ${\tilde {\cal Z}}[m]$ of Eqs.~(\ref{zpartfun-LTfin}) and (\ref{S2Pcurrfin}) is not free, it is still possible to compute exactly the momenta $\langle m^k\rangle$, $k=1,2,\ldots$ of the Gauss linking number:
\begin{equation}
\langle m^k \rangle=\int_{-\infty}^{+\infty} dm \,\,m^k{\tilde {\cal Z}}[m] 
\end{equation}
Indeed, after performing the easy integration over $m$, we obtain:
\begin{eqnarray}
  \langle m^k\rangle &=&\lim_{n_1,n_2\to 0}\lim_{\epsilon_1,\epsilon_2\to 0}\int
        \prod_{i=1}^2\prod_{a_i=1}^{n_i}\left(\int{\cal D}\psi_{i,a_1}^*(\boldsymbol x)
           {\cal D}\psi_{i,a_i}(\boldsymbol x)\right)\nonumber\\
&\times&  \left(\frac 1{4\pi}\int d^3x\epsilon^{\mu\nu\rho}
           a_\mu^{1(cl)}\partial_\nu a_\rho^{2(cl)}\right)^k
           \prod_{i=1}^2|\psi_{i,1}(r_{i,0})|^2|
           e^{-{\tilde S}_2^{\prime\prime}}  \label{mk-LTfin}
\end{eqnarray}
where the fields $a_\mu^{i(cl)}(\boldsymbol x)$ and the action $S^{\prime\prime}_2$ are provided respectively by Eqs.~(\ref{airhosol1}) and (\ref{S2Pcurrfin})
%% appendix sections are then done as normal section
\section{Conclusions}
One of the advantages of the mapping of the polymer problem to field theories is that the partition function of the system and the expectation values of its observables become drastically simpler. This simplification is well known in polymer physics starting from the pioneering works of Edwards and de~Gennes \cite{Edwards,deGennes,desCloizeaux} and is particularly helpful in the presence of topological relations. As a matter of fact, the statistical sum over the polymer paths is intrinsically complicated due to the mathematical complexity of the topological invariants which consist of functionals that are both non-polynomial and non-local in the position vectors $\boldsymbol r_1(s),\ldots,\boldsymbol r_N(s)$.
The examples worked out so far \cite{FFIL1998,non-semisimple,nova,FFHKILPLA2000}, 
show indeed that the partition function of topological constrained polymers can be expressed via local and polynomial field theories. Despite these progresses, a full formulation of the statistical mechanics of polymers in terms of field theoretical models has stumbled against the difficulties arising when attempting to fix the topological constraints with sophisticated topological invariants.

The strategy proposed here to overcome these difficulties is an algorithm to construct field theoretical models that is valid for any numerical topological invariant  $\tau_N(C_1,\ldots,C_N)$ describing the topological states of $N$ loops $C_1,\ldots,C_N$ and with no path ordering. For the first time it has also been possible to include the averages of observables given in the form of Eq.~(\ref{obsgenform}).
The algorithm consists of three simple steps:
\begin{enumerate}
\item Substitute in $\tau_N(C_1,\ldots,C_N)$ all the occurrences of the contour integrals $\oint_{C_i}d\boldsymbol r_i(\cdots)$ over the loops $C_1,\ldots,C_N$ with volume integrals as follows:
\begin{equation}
\oint_{C_i}d\boldsymbol r_i(s)(\cdots)\longrightarrow \int d^3x\boldsymbol b^i(\boldsymbol x)(\cdots)
\end{equation}    
where $\boldsymbol b^i(\boldsymbol x)$ is a magnetic flux density such that:
\begin{equation}
\boldsymbol b^i(\boldsymbol x)=\oint_{C_i} d\boldsymbol r_i(s)\delta(\boldsymbol x - \boldsymbol r_i(s))\label{constroverthebi}
\end{equation}
In the integrand $(\cdots)$ the position vector $\boldsymbol r_i(s)$ should be replaced by the space coordinate $\boldsymbol x$. After these substitutions the topological invariant $\tau_N(C_1,\ldots,C_N)$ does no longer depend on the loop paths $C_1,\ldots,C_N$ and a new quantity $\tau_N(\boldsymbol b^1,\ldots,\boldsymbol b^N)$  is obtained.
The constraints (\ref{constroverthebi}) ensure that $\tau_N(C_1,\ldots,C_N)=\tau_N(\boldsymbol b^1,\ldots,\boldsymbol b^N)$. Examples of this procedure are Eqs.~(\ref{chidblintspat})--(\ref{bydef}) in the case of the Gauss linking number and Eqs.~(\ref{tau3cinb})--(\ref{bi3}) in the case of the Massey invariant (\ref{massey}).
\item In the partition function of the system:
\begin{equation}
  {\cal Z}[m]=\int {\cal D}\boldsymbol r_1\ldots{\cal D}\boldsymbol r_N e^{\sum_{i=1}^N\frac3{2a}{\boldsymbol{\dot r}_i^2(s)}}
  \delta(\tau_N(C_1,\ldots,C_N)-m)
\end{equation}
substitute $\tau_N(C_1,\ldots,C_N)$ with $\tau_N(\boldsymbol b^1,\ldots,\boldsymbol b^N)$. Impose the constraints (\ref{constroverthebi}) on the fields $\boldsymbol b_i(\boldsymbol x)$ using functional Dirac delta functions:
\begin{eqnarray}
  {\cal Z}[m]&=&\int {\cal D}\boldsymbol r_1\ldots{\cal D}\boldsymbol r_N
  \int {\cal D}\boldsymbol b^1\ldots{\cal D}\boldsymbol b^N  e^{\sum_{i=1}^N\frac3{2a}{\boldsymbol{\dot r}_i^2(s)}}
  \delta(\tau_N(C_1,\ldots,C_N)-m)\nonumber\\
  &\times&
  \prod_{i=1}^N\delta\left(
\boldsymbol b^i(\boldsymbol x)-\oint_{C_i}d\boldsymbol r_i(s)\delta(\boldsymbol x-\boldsymbol r_i(s))
  \right)
\end{eqnarray}
In the above equation the dependence on the position vectors $\boldsymbol r_i(s)$ is concentrated in the polymer action and in the functional delta functions. Rewriting the latter using the Fourier representation, the polymer action factorizes into $N$ independent actions describing the fluctuations of a single polymer loop immersed in a magnetic field.
\item  Apply Eq.~(\ref{formbasic}) or alternatively Eq.~(\ref{app5}) in order to eliminate the integrations over the polymer paths and to pass to a local and polynomial field theory.
\end{enumerate}
The averages $\langle {\cal O}\rangle$ of general observables $\cal O$ require an additional recipe:
\begin{description}
\item[\rm 1-bis)] In the expression of an observable $\cal O$ describing the properties of $n$ loops substitute the integrals over the arc-lengths of the $i-$th loop $\int_0^L(\cdots)$ with the volume integrals $\int d^3 x\phi(\boldsymbol x)(\cdots)$, where the fields $\phi_i(\boldsymbol x)$ for $i=1,\ldots,n$ satisfy the conditions:
  \begin{equation}
\phi_i(\boldsymbol x)=\int_0^Lds_i\delta(\boldsymbol x-\boldsymbol r_i(s_i))
  \end{equation}
  Implement these conditions in $\langle {\cal O}\rangle$ by inserting in the related path integral functional Dirac delta functions. Similarly as in step 2), also in the present case it is possible to split the integrations over the paths $\boldsymbol r_i(s)$ into $N$ path integrals in which the sum over these paths can be performed separately using the formulas in Eq.~(\ref{formbasic}) or (\ref{app5}). The fields $\phi_i(\boldsymbol x)$ may be easily integrated out and the computation of  $\langle {\cal O}\rangle$ becomes equivalent to the evaluation of an amplitude of a local and polynomial complex scalar field theory interacting with topological fields.
\end{description}
The above algorithm has been worked out in full details in the case of two different linking invariants, the Gauss linking number
of Eq.~(\ref{glnexplicit}) and the Massey invariant of Eq.~(\ref{massey}). 
The final field theoretical models are respectively given in Eq.~(\ref{zpartfunfin}) and (\ref{Ofinal}).
The initial conditions on the fluctuations of the polymer paths transform into new conditions imposed on the field configurations. As a consequence, the connection with the original topological constraints is not lost after the mapping from polymer paths to fields as it happens in the abelian BF theory coupled to replica scalar fields of Ref.~\cite{FFIL1998}. 
For instance, in the partition function (\ref{zpartfunfin}) a constraint 
is imposed on the cross-helicity of the magnetic fields $\boldsymbol b^1$ and $\boldsymbol b^2$, see Eq.~(\ref{helcond}).
In Eq.~(\ref{Ofinal}) the third-order topological invariant (\ref{thirdorderinv}) restricts the fluxes of the fields $\boldsymbol a^1,\boldsymbol a^2,\boldsymbol a^3$.
Both these invariants are of interest not only for polymer physics, but also for hydrodynamic studies of the solar magnetosphere \cite{Moffatt,Arnold,Hornig}. The Gauss linking number and the triple link invariant $\tau_3(C_1,C_2,C_3)$ of Eq.~(\ref{massey}) are just a few examples of topological invariants that can be included in modeling polymer systems with the approach discussed in this paper.  By applying the same approach to more complicated invariants it will be possible to extract further topological relations between fields. Such relations could be relevant in the investigations of different systems in which the behavior of long and thin filaments is influenced by topology.

Let us note that the number of abelian Chern-Simons fields needed using the prescriptions 1)--3) and 1-bis) in order to take into account the topological constraints imposed by an $N-$th-order invariant $\tau_N(C_1,\ldots,C_N)$ scales as $2N$. This is an important simplification with respect to Ref.~\cite{non-semisimple}, in which six non-Abelian Chern-Simon fields are necessary in order to reproduce the exponent $e^{-i\lambda \tau_3(C_1,C_2,C_3)}$, where
$\tau_3(C_1,C_2,C_3)$ is the Massey type invariant of Eq.~(\ref{massey}).
Moreover, in the present approach the introduction of non-local gauge invariant and metric independent operators like those of Ref.~\cite{non-semisimple} is not required. Only in the case of the Gauss linking number the partition function (\ref{zpartfunfin}) contains an excess of abelian topological fields with respect to the previous model of Ref.~\cite{FFIL1998}. However, we have seen in Subsection~\ref{sub1} that the number of fields can be reduced to two at the price of having to deal with the Fourier parameter $\lambda$. In general, the passage from the fixed value $m$ of the applied numerical invariant to its Fourier conjugate variable $\lambda$ is always possible and is actually helpful in order to get rid of the presence of cumbersome Dirac delta functions.  The disadvantage of that procedure is that  the Fourier parameter $\lambda$ takes values over the whole real line, so that many approximation schemes cannot be applied. Moreover, after any calculation there is the additional difficulty of performing the inverse Fourier transform of the obtained result. 

To check the compatibility of the models builded following the prescriptions 1)--3) and 1-bis) listed above,
in Subsection~\ref{sub1} the partition function of two linked polymers derived in Ref.~\cite{FFIL1998} has been recovered starting from the partition function (\ref{zpartfunfin}).
We have also shown that the theories of topologically entangled polymer rings obtained here
provide several advantages with respect to the previous ones.
For instance, they allow a very efficient calculation of the moments $\langle m^k\rangle$ of the underlying topological invariants. In Section~\ref{sec4}  the expression of $\langle m^k\rangle$
has been reduced in the case of the Gauss linking number to the amplitude of a free field theory that
may be computed in closed form. So far only an approximated expression of the second topological moment $\langle m^2\rangle$ was available \cite{FFHKILPLA2000}. Another important advance is the possibility of mapping the averages of general observables into the expectation values of fields.
An example is the average  $\langle {\cal O}\rangle$ for the wide class of observables $\cal O$
given in Eq.~(\ref{obsgenform}).

Despite the progress represented by the approach discussed in this work with respect to previous analogous methods used to construct analytic and microscopic models of topologically entangled polymer matter, there are
still open problems and several future developments are possible. First of all, we have seen that
the calculation of the moments of the topological invariants has become much easier and, at least in the case of the Gauss linking number, it can be reduced to the evaluation of the amplitude of a free field theory. The extension of this result to more complicated invariants is still missing, but it should be straightforward.
A tougher challenge is certainly provided by the computability of the averages of general observables $\cal O$. Thanks to the new algorithm proposed here, now these averages can be expressed in the form of amplitudes of replica scalar fields coupled to abelian topological fields. However difficulties in their explicit evaluation could be expected due to the presence of a Dirac delta function in the expectation value $\langle{\cal O}\rangle$.
There are only a few  techniques that have been developed in order to cope with theories containing delta functions, see e.~g. Refs.~\cite{FFNPB2012,Chaichian,Achim}. A perturbative approach should also be possible.
The basis for  a potential perturbative treatment are the
classical solutions of Eq.~(\ref{Asolns}) for the fields $\boldsymbol a^i(\boldsymbol x)$  derived
in Subsection~\ref{sub3}. In this way small fluctuations of the polymer rings around a fixed link conformation satisfying the condition (\ref{topconstrFTversion}) could be investigated. Unfortunately, the equations that determine the classical configurations of the replica complex fields and the other couple of topological fields $\boldsymbol c^i(\boldsymbol x)$ are too complicated to be solved in closed form.
In conclusion, it is very likely that new, more powerful methods and approximations will be needed in order to calculate amplitudes such as that in Eq.~(\ref{Ofinal}).
Finally, the algorithm 1)...3) together with step 1-bis) is valid only for numerical topological invariants with no path ordering in the integrations over the contours of the polymer paths. The presence of path ordering in numerical topological invariant is very common. For example the simplest numerical invariant that is able to distinguish the topological states of a single polymer knot contains path ordered contour integrals. This fact has prevented so far the derivation of an analytic model of polymer knots. For this reason it would be highly desirable to generalize the present approach to include invariants with path ordering.

\section{Acknowledgements}
The author would like to acknowledge the contribution of the COST Action CA17139.‌
\appendix
\section{Field theory representation of the Green function of a particle moving inside a magnetic field}\label{appendixa}
The main task of this Appendix is the derivation of Eq.~(\ref{formbasic}).
The starting point is the Lagrangian (\ref{lagrbasic}) that will be rewritten here for convenience in the form:
\begin{equation}
  {\cal L}(\boldsymbol {\dot r}(s),\boldsymbol r(s),\boldsymbol A)=\frac 3{2a}\boldsymbol{\dot r}^2(s)+i\boldsymbol {\dot r}(s) \cdot \boldsymbol A(\boldsymbol r(s))+i\eta(\boldsymbol r(s))\label{lagrbasicapp}
\end{equation}
Apart from the presence of the complex factor $i$ in the magnetic potential, this Lagrangian can be interpreted as that of a particle immersed in a magnetic field.
The related Hamiltonian is:
\begin{equation}
H=\frac a6\left( \boldsymbol p-i\boldsymbol A\right)^2-i\eta(\boldsymbol r)
\end{equation}
First, we suppose that the polymer is an open chain with ends located in the fixed points $\boldsymbol r$ and $\boldsymbol r_0$. The case of a polymer ring will be obtained in the limit:
\begin{equation}
\boldsymbol r(L)=\boldsymbol r=\boldsymbol r(0)=\boldsymbol r_0\label{ringlimit}
\end{equation}
The path integral
\begin{equation}
{\cal I}[\boldsymbol r,\boldsymbol r_0]=\int_{\boldsymbol {r}(0)=\boldsymbol r_0}^{\boldsymbol r(L)=\boldsymbol r}{\cal D}\boldsymbol r_i(s) e^{-{\cal L}(\boldsymbol {\dot r}_i(s),\boldsymbol r_i(s),\boldsymbol A)}\label{app0}
\end{equation}
is the matrix element between the states $\langle \boldsymbol r|$ and $|\boldsymbol r_0\rangle$ of the evolution operator $e^{-HL}$:
\begin{equation}
{{\cal I}[\boldsymbol r,\boldsymbol r_0]}=\langle \boldsymbol r|e^{-HL}|\boldsymbol r_0\rangle
\end{equation}
$\langle \boldsymbol r|e^{-HL}|\boldsymbol r_0\rangle$ satisfies the equation:
\begin{equation}
  \left[
\frac \partial{\partial L}-\frac a6\left(\boldsymbol \nabla-i\boldsymbol A\right)^2 -i\eta
    \right]\langle \boldsymbol r|e^{-HL}|\boldsymbol r_0\rangle=\delta(L)\delta(\boldsymbol r-\boldsymbol r_0)\label{app1}
\end{equation}
together with  the boundary condition:
\begin{equation}
\left.\langle \boldsymbol r|e^{-HL}|\boldsymbol r_0\rangle\right|_{L=0}=\delta(\boldsymbol r-\boldsymbol r_0)
\end{equation}
From Eq.~(\ref{app1}) it turns out that $\langle \boldsymbol r|e^{-HL}|\boldsymbol r_0\rangle$ is the inverse of the operator $ \left[
\frac \partial{\partial L}-\frac a6\left(\boldsymbol \nabla-i\boldsymbol A\right)^2-i\eta
\right]$ and can be written as the Green function of a complex scalar field theory as follows:
\begin{equation}
  .\langle \boldsymbol r|e^{-HL}|\boldsymbol r_0\rangle=\frac{\int {\cal D}\psi^*{\cal D}\psi e^{\int d^3x\int_0^Ldt\psi^*\left[\frac\partial{\partial t}-\frac a6\left(\boldsymbol \nabla-i\boldsymbol A\right)^2-i\eta\right]\psi}\psi^*(\boldsymbol r,L)\psi(\boldsymbol r_0,0)}{
\int {\cal D}\psi^*{\cal D}\psi e^{\int d^3x\int_0^Ldt\psi^*\left[\frac\partial{\partial t}-\frac a6\left(\boldsymbol \nabla-i\boldsymbol A\right)^2-i\eta\right]\psi}
  }\label{app2}
\end{equation}
Finally, to get rid of the denominator in the right hand side of the above equation, we use the replica trick $Z^{-1}= \lim_{n\to 0}Z^{n-1}$. To this purpose we introduce a set of $n$ replica complex fields $\psi_a^*,\psi_a$, $a=1,\ldots,n$. In this way, putting together Eqs.~(\ref{app0})--(\ref{app2}), it is possible to conclude that
\begin{eqnarray}
  \int_{\boldsymbol {r}(0)=\boldsymbol r_0}^{\boldsymbol r(L)=\boldsymbol r}{\cal D}\boldsymbol r_i(s) e^{-{\cal L}(\boldsymbol {\dot r}_i(s),\boldsymbol r_i(s),\boldsymbol A)}&=&\lim_{n\to 0}\int\left[\prod_{a=1}^n{\cal D}\psi_a^*{\cal D}\psi_a
    \right]\psi^*_{1}(\boldsymbol r,L)\psi_1(\boldsymbol r_0,0)\nonumber\\
  &\times&e^{\sum\limits_{a=1}^n\int d^3x\int_0^Ldt \psi^*_a\left[
      \frac\partial{\partial t}-\frac a6\left(
\boldsymbol \nabla-i\boldsymbol A
      \right)^2  -i\eta  \right]
\psi_a
    }
\end{eqnarray}
In the limit of a closed ring (\ref{ringlimit}) in which the point $\boldsymbol r(L)=\boldsymbol r(0)=\boldsymbol r_0$, the above equation becomes:
\begin{eqnarray}
  \oint_{\boldsymbol {r}(0)=\boldsymbol r_0}{\cal D}\boldsymbol r_i(s) e^{-{\cal L}(\boldsymbol {\dot r}_i(s),\boldsymbol r_i(s),\boldsymbol A)}&=&\lim_{n\to 0}\int\left[\prod_{a=1}^n{\cal D}\psi_a^*{\cal D}\psi_a
    \right]|\psi_{1}(\boldsymbol r_0,0)|^2\nonumber\\
  &\times&e^{\sum\limits_{a=1}^n\int d^3x\int_0^Ldt \psi^*_a\left[
      \frac\partial{\partial t}-\frac a6\left(
\boldsymbol \nabla-i\boldsymbol A
      \right)^2 -i\eta   \right]
\psi_a}
\end{eqnarray}
This completes the derivation of Eq.~(\ref{formbasic}).

An useful variant of this formula can be obtained taking the Laplace transform of $\langle \boldsymbol r|e^{-HL}|\boldsymbol r_0\rangle$:
\begin{equation}
\int_0^{+\infty}e^{-zL}\langle \boldsymbol r|e^{-HL}|\boldsymbol r_0\rangle = \langle \boldsymbol r|\frac1{z+H}|\boldsymbol r_0\rangle\label{app3}
\end{equation}
with $z\ge0$. The Green function $\langle \boldsymbol r|\frac1{z+H}|\boldsymbol r_0\rangle$ is the inverse of the operator $z+H$.
Using a similar strategy as we did in the case of $\langle \boldsymbol r|e^{-HL}|\boldsymbol r_0\rangle$, it is possible to express also $\langle \boldsymbol r|\frac1{z+H}|\boldsymbol r_0\rangle$ in the form of the propagator of complex replica fields:
\begin{eqnarray}
\langle \boldsymbol r_0|\frac1{z+H}|\boldsymbol r_0\rangle&=&-\frac{\delta^2}{\partial J_1(\boldsymbol r)\delta J^*_1(\boldsymbol r_0)}\lim_{n\to 0}\lim_{\epsilon\to 0}\int\left[\prod_{a=1}^n{\cal D}\psi^*_a{\cal D}\psi_a
    \right]e^{-\epsilon\int d^3x\psi^*_a\psi_a}\nonumber\\
  &\times&\left. e^{i\sum\limits_{a=1}^n\int d^3x \psi^*_a\left[
      (z+i\epsilon)-\frac a6\left(
\boldsymbol \nabla-i\boldsymbol A
      \right)^2 -i\eta   \right]
\psi_a
    +J_a^*\psi_a+J_a\psi^*_a}\right|_{J_a^*=J_a=0}\label{app4}
\end{eqnarray}
Let us notice in the above equation the presence of an overall purely imaginary $i$ factor in the action and the term $-\int d^3x \psi_a^*\psi_a$ to guarantee convergence. Performing in the above equation the functional derivatives with respect to $J^*_1(\boldsymbol r_0)$ and $J_1(\boldsymbol r)$ we arrive to the final result:
\begin{eqnarray}
\langle \boldsymbol r_0|\frac1{z+H}|\boldsymbol r_0\rangle&=&\lim_{n\to 0}\lim_{\epsilon\to 0}\int\left[\prod_{a=1}^n{\cal D}\psi^*_a{\cal D}\psi_a
    \right]|\psi_1(\boldsymbol r_0)|^2\nonumber\\
  &\times&e^{i\sum\limits_{a=1}^n\int d^3x \psi^*_a\left[
      (z+i\epsilon)-\frac a6\left(
\boldsymbol \nabla-i\boldsymbol A
      \right)^2  -i\eta    \right]
\psi_a
}\label{app5}
\end{eqnarray}



\end{document}